\documentclass{aa}
\usepackage{graphicx}
\usepackage{amssymb}
\usepackage{txfonts}
\newcommand{\kms}{km\,s$^{-1}$}

\newcommand{\Msol}{M$_{\odot}$}

\newcommand{\Tex}{$T_{\rm ex}$}
\newcommand{\Tmb}{$T_{\rm mb}$}
\newcommand{\TAstar}{$T_{\rm A}^*$}
\newcommand{\ITAstar}{$\int T_{\rm A}^* {\rm dv}$}
\newcommand{\ITmb}{$\int T_{\rm mb} {\rm dv}$}
\newcommand{\Vlsr}{$V_{\rm lsr}$}
\newcommand{\amin}{$^{\prime}$}
\newcommand{\asec}{$^{\prime\prime}$}
\newcommand{\pad}{.\hskip-2pt$^\circ$}
\newcommand{\pam}{.\hskip-2pt$^{\prime}$}
\newcommand{\pas}{.\hskip-2pt$^{\prime\prime}$}

\begin{document}
\title{Molecular gas and stars in the translucent cloud MBM~18 (LDN~1569)
\thanks{Partly based on observations collected at the European Southern 
Observatory (ESO) using the ESO 3.6-m and the Swedish-ESO Submillimetre 
Telescope (SEST), La Silla, Chile},
\thanks{Appendix A is available in electronic form at http://www.aanda.org},
\thanks{The full spectra are available at the CDS via anonymous ftp to 
cdsarc.u-strasbg.fr (130.79.128.5) or via 
http://cdsweb.u-strasbg.fr/viz-bin/qcat?J/A+A/vol/page}
}


\author{J. Brand\inst{1}
          \and
          J. G. A. Wouterloot\inst{2}
          \and
          L. Magnani\inst{3}
          }

\offprints{J. Brand}

\institute{INAF - Istituto di Radioastronomia \& Italian ALMA Regional Centre, 
      Via P. Gobetti 101, 40129 Bologna, Italy\\
    \email{brand@ira.inaf.it}
    \and Joint Astronomy Centre,
         660 N. A'Ohoku Place, University Park, Hilo, HI 96720, USA \\
     \email{j.wouterloot@jach.hawaii.edu}
     \and Department of Physics and Astronomy, University of Georgia, 
          Athens, GA 30602, USA\\
     \email{loris@physast.athens.edu}
             }

\date{ }

\abstract{We investigate star formation in translucent, high-latitude clouds.}
{Our aim is to understand the star-formation history and rate in the solar 
neighbourhood.}
{We used spectroscopic observations of newly found candidate H$\alpha$ 
emission-line stars to 
establish their pre-main-sequence nature. The environment was studied through 
molecular line observations of the cloud (MBM~18/LDN~1569) in which the stars 
are presumably embedded.}
{Ten candidate H$\alpha$ emission-line stars were found in an objective grism 
survey of a $\sim 1$~square degree region in MBM~18, of which seven have been 
observed spectroscopically in this study. Four of these have weak 
($\mid$ W(H$\alpha) \mid \la 5$~\AA) H$\alpha$ emission, and six 
out of seven have spectral types M1$-$M4~V. One star is  
of type F7-G1~V, and has H$\alpha$ in absorption. 
The spectra of three of the M-stars may show an absorption line of LiI, 
although none 
of these is an unambiguous detection. The M-stars lie at distances between 
$\sim$60~pc and 250~pc, while most distance determinations of MBM~18 
found in the literature agree on $120-150$~pc.
For the six M-stars a good fit is obtained with pre-main-sequence 
isochrones indicating ages between 7.5 and 15~Myr.
The mass of the molecular material, derived from the integrated 
$^{12}$CO(1--0) emission, is $\sim$ 160~\Msol\ (for a distance of 120~pc). 
This is much smaller than the virial mass ($\sim 10^3$~\Msol), and the 
cloud is not gravitationally bound. Using a clump-finding routine, we identify 
12 clumps from the CO-data, with masses between 2.2 and 22~\Msol. All clumps  
have a virial mass at least six times higher than their CO-mass, and thus none 
are in gravitational equilibrium. A similar situation is found from 
higher-resolution CO-observations of the northern part of the cloud.}
{Considering the relative weakness or absence of the H$\alpha$ emission, the 
absence of other emission lines, and the lack of clear LiI absorption, 
the targets are not T Tauri stars.
With ages between 7.5 and 15~Myr they are old enough to explain the lack of 
lithium in their spectra. 
Based 
on the derived distances, some of the stars may lie inside the molecular 
cloud. From the fact that the cloud as a whole, as well as the individual 
clumps, are not gravitationally bound, in combination with the 
ages of the stars we conclude that it is 
not likely that (these) stars were formed in MBM~18. 
}


\maketitle
%

\section{Introduction}

\begin{figure*}
\resizebox{18cm}{!}{\includegraphics{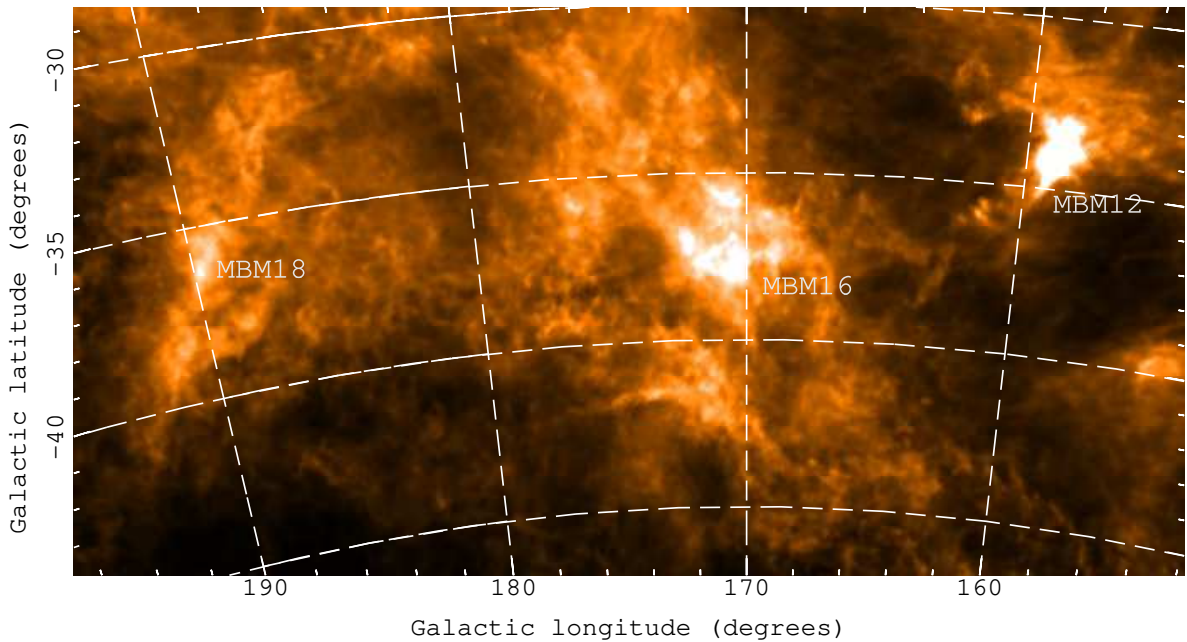}}
\caption{Map of the colour excess E(B$-$V) from Schlegel et al.~\cite{schlegel}.
The map is centred at {\sl l} $\approx$ 174\degr, {\sl b}=$-37$\degr\ 
and covers roughly 40\degr\ $\times$ 15\degr; a grid in galactic 
coordinates has been superposed. The minimum and maximum values in the map 
are 0.03 and 2.09~magnitudes, respectively.
The 3 principal features in this map from left to right are LDN~1569 (MBM~18),
MBM~16 slightly to the right of the centre in the map, and MBM~12 
(LDN~1453/4/7/8). 
MBM~12 is a known star-forming region with a 
young T-association that includes about a dozen objects, MBM~16 is not known 
to be forming stars, and LDN~1569 (MBM~18)
contains the possible pre-main-sequence stars discussed in the text.
}
\label{dustmap}
\end{figure*}

Star formation in the Galaxy proceeds fitfully in both small and giant
molecular clouds. The general mechanism (external or internal
triggers leading to the formation of single stars or clusters) has been
identified, though many details still must be worked out (see, e.g.,
Stahler \& Palla~\cite{pallabook}). Among the 
lower-mass clouds, the star-formation
capability of translucent clouds (objects with $1<A_{\mathrm V}<5$~mag) 
has not been well-determined yet. Some clouds near the translucent/dark cloud
demarcation clearly form stars (e.g., MBM~12\footnote{The MBM-identifier 
comes from the original catalogue of Magnani, 
Blitz \& Mundy (\cite{mbm}).}, also known as LDN~1453/4/7/8; 
Luhman~\cite{luhman}), while other, somewhat similar objects, do not (MBM~40; 
Magnani et al.~\cite{magnani96a}). The
reasons for this marked difference in star-forming capability are not
clear. Unfortunately, the samples of star-forming
vs. non-star-forming translucent clouds are so small that it is
difficult to recognize patterns or trends. While more than a hundred translucent
clouds are known to exist (the vast majority
identified at high Galactic latitudes -- see Magnani et 
al.~\cite{magnani96b}), less than a dozen have been thoroughly
searched for evidence of star formation. Of these, only MBM~12 and MBM~20 
(LDN~1642) show unambiguous evidence of low-mass star formation. 
About a half-dozen other clouds have some indication of possible star
formation; see review by McGehee (\cite{mcgehee}). 
The question of whether or
not stars form in translucent clouds is important because of the
transient nature of these objects. Most translucent clouds at high
Galactic latitudes are not gravitationally bound and are breaking up
on timescales of 10$^6$ years (Magnani et al.~\cite{mbm}). 
While a given translucent cloud may be gravitationally unbound, individual 
clumps within the cloud can be bound and potentially host star formation. 
Thus, the mechanism by which low-mass
star formation begins in these objects is likely to differ from the
mechanisms invoked for the larger, denser, dark molecular clouds.
Determining which translucent high-latitude clouds do form stars and
which do not could potentially unravel the nature of this process and shed 
more light on star formation in all low-mass molecular
clouds. To accomplish this objective,
it is imperative to expand the list of translucent clouds that
have been thoroughly searched for low-mass star formation.

\bigskip
A distinguishing characteristic of low-mass pre-main-sequence (T Tauri) stars 
is the presence of emission lines in their spectra, 
especially H$\alpha$. Searches for T Tauri stars are often carried out 
through H$\alpha$ objective grism surveys. Stars found in this way have to 
be studied spectroscopically, to confirm their pre-main-sequence nature (and 
to rule out, e.g., dMe or Be stars).
We have carried out an objective prism search for H$\alpha$ emission-line 
stars in various types of clouds with the ESO 3.6-m telescope 
(Sect.~\ref{obs3p6}). One of our targets was 
the high-latitude translucent cloud 
\object{LDN~1569} (\object{MBM~18}; {\sl b}=$-35$\degr). 
Located at a distance of 
$\sim$120--150~pc (Penprase~\cite{penprase93}; but see Sect.~\ref{discus}), 
15\degr\ southeast 
of the Taurus/Auriga dark cloud complex, the cloud is of similar size and
structure to MBM~12, but has lower extinction by a factor of 3 (at a 
resolution of 6\pam1; see Fig.~\ref{dustmap}).
We have found ten candidate H$\alpha$ emission-line 
stars in a 45\arcmin\ $\times$ 45\arcmin\ area centred on this cloud.

\bigskip
To establish whether our candidates are T Tauri stars requires 
higher-resolution spectra to seek confirmation of 
the presence of H$\alpha$ emission and to determine their  
spectral types. The presence of LiI absorption at 6708~\AA\ above a 
certain equivalent width threshold in the spectra (see e.g. Preibisch et 
al.~\cite{preibisch}; Bertout~\cite{bertout}) would also give an indication of 
the stars' youth (cf. Luhman~\cite{luhman}). It also needs to be established 
whether the stars are associated with the cloud against which they are seen 
projected. 
Determining that the candidates from the objective grism survey are 
truly pre-main-sequence stars formed in MBM~18 would be an important
result because it would extend the extinction range of those
translucent clouds capable of forming stars down to objects with $\sim 2$ 
magnitudes of visual extinction in their densest regions. There are currently 
no unambiguous signatures of star formation in any high-latitude cloud with 
$A_{\rm V} < 4$~mag in its most opaque region. Moreover, the cloud
is likely at 100--200~pc from the Sun, thus the pre-main-sequence stars
would be among the nearest such objects, and if the objects
constitute a small association, this would be a significant finding,
on a par with the discovery of the TW Hydrae association (e.g., de la Reza et 
al.~\cite{delareza}; Zuckerman et al.~\cite{zuckerman}, and references 
therein), and would
reveal important information on the star-formation history and rate in
the solar neighbourhood.  

We also need to obtain information on the environment of the stars, in 
particular the dynamical state of the gas (gravitationally bound or not) of 
the cloud (and its substructures) towards which the stars are seen 
projected.

In this paper we present the results 
of higher-resolution spectroscopy of emission-line star candidates in MBM~18, 
and of molecular line observations of the gas in this cloud.

\begin{table*}
\caption{Coordinates and magnitudes of the candidate H$\alpha$ emission-line
stars, and log of TNG observations.}
\label{targets}
\begin{flushleft}
\begin{tabular}{cccccccccccccll}
\hline
\noalign{\smallskip}
Star & \multicolumn{6}{c}{RA\ \ (J2000)\ \ DEC} & F$^a$ & V$^a$ & J$^b$ & H$^b$ & K$^b$ & TNG & int. time & comment \\
   & h & m & s & $\circ$ & $\prime$ & $\prime\prime$ & \multicolumn{5}{c}{mag} & dd/mm/yy & sec & \\
\noalign{\smallskip}
\hline
\noalign{\smallskip}
Ha1 & 04 & 00 & 42.17 & +00 & 45 & 09.9 & 14.90 & 15.69 & 12.10 & 11.50 & 11.26 & 14/11/10 & $4 \times 1728$ & \\
Ha2 & 04 & 00 & 43.13 & +00 & 53 & 17.4 & 15.06 & 15.96 & 11.92 & 11.27 & 11.03 & 15/11/10 & $2 \times 1440$ & \\
Ha3 & 04 & 00 & 55.40 & +01 & 04 & 40.0 & 16.98 & 17.62 & 14.56 & 13.72 & 13.47 & \multicolumn{2}{c}{not observed} & \\
Ha4 & 04 & 00 & 58.53 & +00 & 44 & 33.8 & 16.15 & 16.81 & 14.86 & 14.37 & 14.35 & 14/11/10 & $4 \times 1728$ & in same slit as Ha1 \\
Ha5 & 04 & 01 & 23.56 & +01 & 06 & 49.1 & 14.57 & 15.51 & 10.89 & 10.32 & 10.00 & 12/03/09 & $4 \times 450$  & \\
Ha6 & 04 & 01 & 24.70 & +01 & 07 & 22.8 & 16.50 & 17.86 & 11.89 & 11.29 & 10.94 & 12/03/09 & $4 \times 450$  & in same slit as Ha5 \\
Ha7 & 04 & 02 & 44.80 & +01 & 13 & 15.5 & 15.89 & 16.60 & 13.36 & 12.78 & 12.54 & 29/10/10 & $4 \times 1620$ & \\
Ha8 & 04 & 03 & 00.07 & +00 & 27 & 35.0 & 15.19 & 15.86 & 12.14 & 11.56 & 11.35 & 12/12/10 & $2 \times 1620$ & \\
Ha9 & 04 & 03 & 30.65 & +00 & 32 & 28.4 & 16.37 & 17.42 & 13.35 & 12.67 & 12.37 & \multicolumn{2}{c}{not observed} & \\
Ha10 & 04 & 03 & 35.75 & +00 & 34 & 13.0 & 17.88 & 18.74 & 13.63 & 13.09 & 12.75 & \multicolumn{2}{c}{not observed} & \\
\noalign{\smallskip}
\hline
\end{tabular}
\smallskip\noindent
\\
$^a$\ Photographic magnitude from the GSC; F-mag approximately corresponding to the Johnson R-mag.\\
$^b$\ Infrared magnitudes from the 2MASS survey. \\
\end{flushleft}
\end{table*}

\begin{figure*}
\resizebox{18cm}{!}{\includegraphics{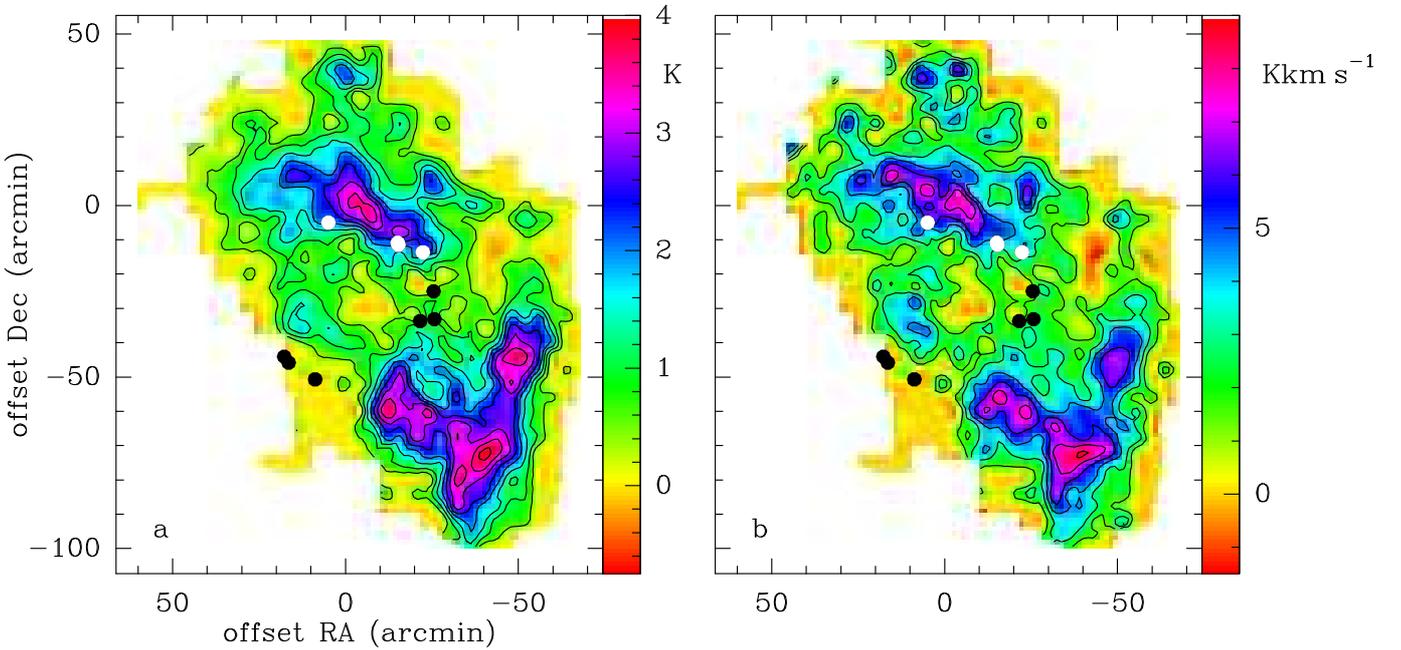}}
\caption{
{\bf a.}\ Distribution of peak $T_\mathrm{A}^*$ of $^{12}$CO(1--0) towards 
LDN~1569 obtained from KOSMA observations and Gaussian fits to the 
spectra. Contour levels are 0.5 -- 4~K in steps of 0.5~K. 
{\bf b.}\ as a., but for the line-integrated emission \ITAstar.
Contour levels are 1.5 -- 9.0~Kkm\,s$^{-1}$ in steps of 1.5~Kkm\,s$^{-1}$. 
Offsets are in arcminutes relative to 
4$^h$02$^m$25.0$^s$, 01$^{\circ}$18$^{\prime}$17$^{\prime\prime}$ (J2000).
The filled circles indicate the positions of the candidate emission-line stars.
}
\label{kosma_tp_ti}
\end{figure*}

\section{Observations}

\subsection{ESO 3.6-m \label{obs3p6}}

On 21 November 1982 we used the ESO 3.6-m telescope at La Silla, in 
combination with the triplet corrector and a 1540~\AA mm$^{-1}$ red-sensitive 
grism in its prime focus, to obtain a plate with
low-resolution spectra of a 1 square degree area of LDN~1569, covering the 
wavelength range 6300 -- 9000~\AA. This range is determined by the use of a 
IV-N plate 
and an RG630 filter. Just prior to the observations the plate was sensitised 
in AgNO$_3$, and later developed in D19.
The field centre was 
4$^h$02$^m$04.5$^s$, 00$^{\circ}$48$^{\prime}$18$^{\prime\prime}$ (J2000)
(at offset $-$5\pam1,$-$30\pam0 with respect to the (0,0) reference position 
in the CO maps). The limiting red magnitude was about 21.
In the same run we obtained a similar plate of LDN~1641 in Orion (not a 
translucent cloud; Wouterloot \& Brand~\cite{wb92}; see their Fig.~1 for 
examples of the grism-spectra).

\subsection{TNG}

On 12 March 2009, on 29 October, 14 and 15 November, and on 12 December 2010 
we used the low-resolution spectrograph ({\it DOLORES}) at the Italian 
Telescopio Nazionale Galileo (TNG) at La Palma (Spain) to perform long-slit 
spectroscopy on seven of the candidate H$\alpha$ emission-line 
stars detected in the grism-survey. 
The coordinates of the targets and their estimated magnitudes are listed 
in Table~\ref{targets}, which also presents the log of the TNG observations.

We used grism VHR-R, which covers a wavelength range of 6240--7720~\AA\ with a 
dispersion of 0.80~\AA/pix. The scale of the CCD detector is 0\pas252/pixel. 
The observations were carried out with a slit width of 1\asec\ or 1\pas5, 
depending on the seeing, resulting in a spectral resolution of 3.2~\AA\ and
4.8~\AA, respectively.

To avoid problems with cosmic rays, several separate spectra per star were 
obtained. 
The number of spectra and the individual integration times are listed in 
Table~\ref{targets}.
Two of the stars (Ha4 and Ha6) were observed simultaneously with another 
target (Ha1 and Ha5, respectively) by positioning the slit at 
an appropriate angle. The integration time was based on the brighter 
star in the slit, thus the signal-to-noise ratio for the other 
target is lower than for the primary one.

To allow absolute flux calibration the standard star Feige24 or Feige34 
(for Ha5-Ha6) was observed immediately before or after the target 
observations, using the same instrumental 
setup as for the target observations. Flat-fielding was performed using 
10 (5 for Ha5-Ha6) frames which were uniformly illuminated by a 
halogen lamp. Wavelength 
calibration was performed using an arc-spectrum of an Ar, Ne+Hg, and Kr lamp, 
or a Ne+Hg (for Ha7) comparison lamp. 
A bias frame, to be subtracted from the other frames before analysis, was 
constructed from ten individual bias frames. Flat-, arc-, and bias-frames 
were obtained on the same day as the science observations and with the same 
instrumental setup. 

\begin{figure}
\resizebox{9cm}{!}{\includegraphics{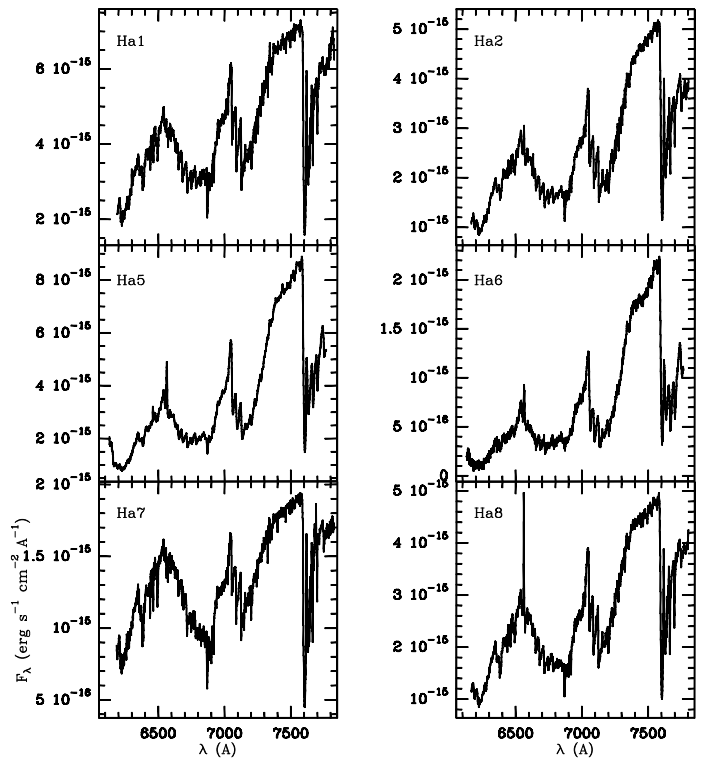}}
\caption{
Overview of the spectra obtained with Dolores at the TNG for the targets 
identified in each panel, showing the characteristic shape 
and absorption bands of M-type dwarf stars. More detailed spectra are 
shown in the appendix (on line version only).
}
\label{thumbnails}
\end{figure}

\subsection{Molecular line observations}

\subsubsection{KOSMA}
We mapped the complete LDN~1569 cloud in $^{12}$CO J=1--0 using the KOSMA 
3-m telescope (Gornergrat, Zermatt, Switzerland) between September 12 
and December 1, 1989. We observed 708 positions on a
4~arcmin raster using frequency switching. 
The frequency resolution of the acousto-optical spectral (AOS) backend used 
was 0.1665 MHz (0.433~\kms). After folding the frequency-switched spectra, 
the resulting rms was $0.16-0.26$~K
($T_\mathrm{A}^*$) with a median value of 0.21~K. The FWHM beam size at 
115~GHz was 3\pam9, and the main-beam efficiency of the telescope was 0.78.

\subsubsection{SEST}
The northern part of LDN~1569 was mapped with the 15-m SEST (ESO, La Silla,
Chile) in $^{12}$CO 
J=1--0 between November 18 and 26, 1988. We observed 1570 positions 
on a 40 or 80 arcsec raster using frequency switching.
We used an AOS with a spectral resolution of 0.043~MHz (0.113~\kms), and
the rms in the spectra after folding was 0.06-0.40~K ($T_\mathrm{A}^*$) 
(median value 0.24~K). The main-beam efficiency at 115~GHz was 0.70. 
The FWHM beamsize of the SEST at this frequency is about 46\arcsec.

The central region of the northern cloud was then mapped (289 positions) 
in $^{13}$CO J=1--0 in the same way. The rms (velocity resolution 
0.118~km\,s$^{-1}$) was 0.03 - 0.13~K ($T_\mathrm{A}^*$; median 0.10~K).
A few positions were observed in C$^{18}$O J=1--0 and CS J=2--1, both with 
rms 0.04~K ($T_\mathrm{A}^*$).

In addition, in January 1990 we observed 54 positions in $^{12}$CO(2--1), mostly
located around offset $-80$\arcsec,$-80$\arcsec\ and along two lines through 
the edge of the cloud, while three positions (at offsets 0\arcsec,0\arcsec, 
$-80$\arcsec,$-80$\arcsec, and $-800$\arcsec,$-400$\arcsec) were observed in 
$^{13}$CO(2--1). 
The velocity resolution was
0.051~km\,s$^{-1}$ and the main-beam efficiency was about 0.5. The rms 
($T_\mathrm{A}^*$; spectra smoothed to 0.113~km\,s$^{-1}$) was $0.15 - 0.40$~K 
for $^{12}$CO(2--1), and $\sim 0.10$~K for $^{13}$CO(2--1), respectively.

\subsubsection{JCMT}
To confirm some results of the KOSMA map (namely that there is only weak CO 
emission at a far-IR maximum) we observed four positions with the
JCMT 15-m telescope (Mauna Kea, Hawaii, USA) in $^{12}$CO J=2--1. The 
observations were carried out on 28 October 
2008 in bad but stable weather. The rms ($T_\mathrm{A}^*$) in those spectra 
was $0.26 - 0.33$~K (resolution 0.040~\kms); a standard observation on 
CRL618 showed the expected intensity. The beam FHWM at this frequency is 
20\arcsec, and the main-beam efficiency was 0.75. The pointing accuracy was 
about 2\arcsec.

\subsubsection{Effelsberg}
Between March 28 and May 4, 1992 we used the 100-m Effelsberg telescope to
observe NH$_3$(1,1) and (2,2) towards the core of the SEST $^{13}$CO cloud.
We observed 26 positions on a 40 arcsec raster with a resolution of
12.177~kHz (0.154~\kms). The rms of most spectra was $0.04 - 0.08$~K (\Tmb).
The beam FWHM at this frequency is 40\arcsec.

\subsubsection{Medicina}
Between $10 - 12$ October 2008, on 2 April and 16 May 2009, and on 19 January, 
20 February, and 18 March 2011 we used the Medicina 32-m telescope\footnote{The 
Medicina 32-m VLBI antenna is operated by INAF--Istituto di Radioastronomia} 
to look for H$_2$O($6_{16}-5_{23}$) maser emission in the direction of all 
candidate H$\alpha$ emission-line stars except stars Ha3 and Ha4.

We used a bandwidth of 10~MHz and 1024 channels, resulting in a resolution 
of 9.77~kHz (0.132\kms); the HPBW at 22.235~GHz was $\sim$1\pam 9.

The telescope 
pointing model is typically updated a few times per year, and is quickly 
checked every few weeks by observing strong maser sources (e.g., W3~OH, 
Orion-KL, W49~N, Sgr~B2, and W51). The pointing accuracy was always better 
than 25\arcsec; the rms residuals from the pointing model were of the order 
of 8\arcsec--10\arcsec. 

Observations were taken in total power mode, with 
both ON and OFF scans of 5~min duration. The OFF position was taken 
1\pad25 E of the source position to rescan the same path as the ON scan. 
Typically, two ON/OFF pairs were taken at each position, though during the 
01/2011 session we took ten pairs on Ha7. 

The antenna gain as a function of elevation was determined by observing 
the continuum source DR~21 several times per day (for which we assume a 
flux density of 16.4 Jy after scaling the value of 17.04~Jy given by  
Ott et al. (\cite{ott}) for the ratio of the source size to the Medicina 
beam) at a range of elevations. Antenna temperatures were derived from 
total power measurements in position-switching mode. The integration time 
at each position was 10 sec with 400~MHz bandwidth. The zenith system 
temperature was about $90-100$~K in clear weather conditions. 

The daily gain curve was determined by fitting a polynomial curve to the 
DR~21 data; this was then used to convert antenna temperature to flux 
density for all spectra taken that day. From the dispersion of the single 
measurements around the curve, we found the typical calibration uncertainty 
to be 20\%.

\section{Data reduction and results}

\subsection{ESO 3.6-m \label{red3p6}}

The plate was scanned with the Astroscan measuring machine at the Leiden 
Observatory (NL) and digitized.
Spectra of all stars on the plate were extracted and analysed with
an algorithm that searches for a peak at the wavelength of H$\alpha$ in the 
first-order spectrum.
These spectra were also inspected visually. In this way we 
obtained a list of ten candidate H$\alpha$ emission-line stars in 
a 45\arcmin\ $\times$ 45\arcmin\ area in LDN~1569 (see Table~\ref{targets}). 
In Fig.~\ref{kosma_tp_ti} 
we show the locations of the stars superimposed on the map of integrated 
CO emission.  

\subsection{TNG - spectra\label{sub-tng}}

Data were reduced with the IRAF package. From all science frames a bias 
was subtracted, after which they were divided by the normalised flat field. 
From each of the science frames the trace(s) of the star(s) were 
extracted and these were wavelength-calibrated using one of the frames 
with the arc-spectrum. Each target was wavelength-calibrated with the 
arc-spectrum extracted at the same location on the detector, to compensate 
for small deviations that might occur in the alignment of the reference 
emission lines across the detector. The spectra were then corrected for 
extinction, and flux-calibrated using the standard star observations. 
The individual one-dimensional wavelength- and flux-calibrated spectra of 
each target were then averaged into a final spectrum. 
To further correct the wavelength calibration, we used 
the sky lines that were subtracted from the stellar spectra. For each 
spectrum, Gaussian fits 
were made to tens of sky lines, and their wavelengths were compared to those 
listed in Osterbrock et al. (\cite{osterbrock}). Three stars were found to need 
a small correction: Ha2 ($-$1.5~\AA) and Ha5 and 6 (both $-2.2$~\AA). For the 
other four stars the difference was negligible, although for the sky lines in 
Ha1 and Ha4 (which were observed in the same slit) the deviation between 
measured and literature wavelengths varied slightly, but systematically, with 
wavelengths between 6250~\AA\ and 7600~\AA, while at longer wavelengths the 
deviations became rapidly larger (up to several Angstroms). 
The final spectra are shown in Fig.~\ref{fullspectra1} in the appendix 
(available in the on line version only). 
Smaller 
versions (except for Ha4, see below) are presented in Fig.~\ref{thumbnails}

\begin{figure*}
\resizebox{17cm}{!}{\includegraphics{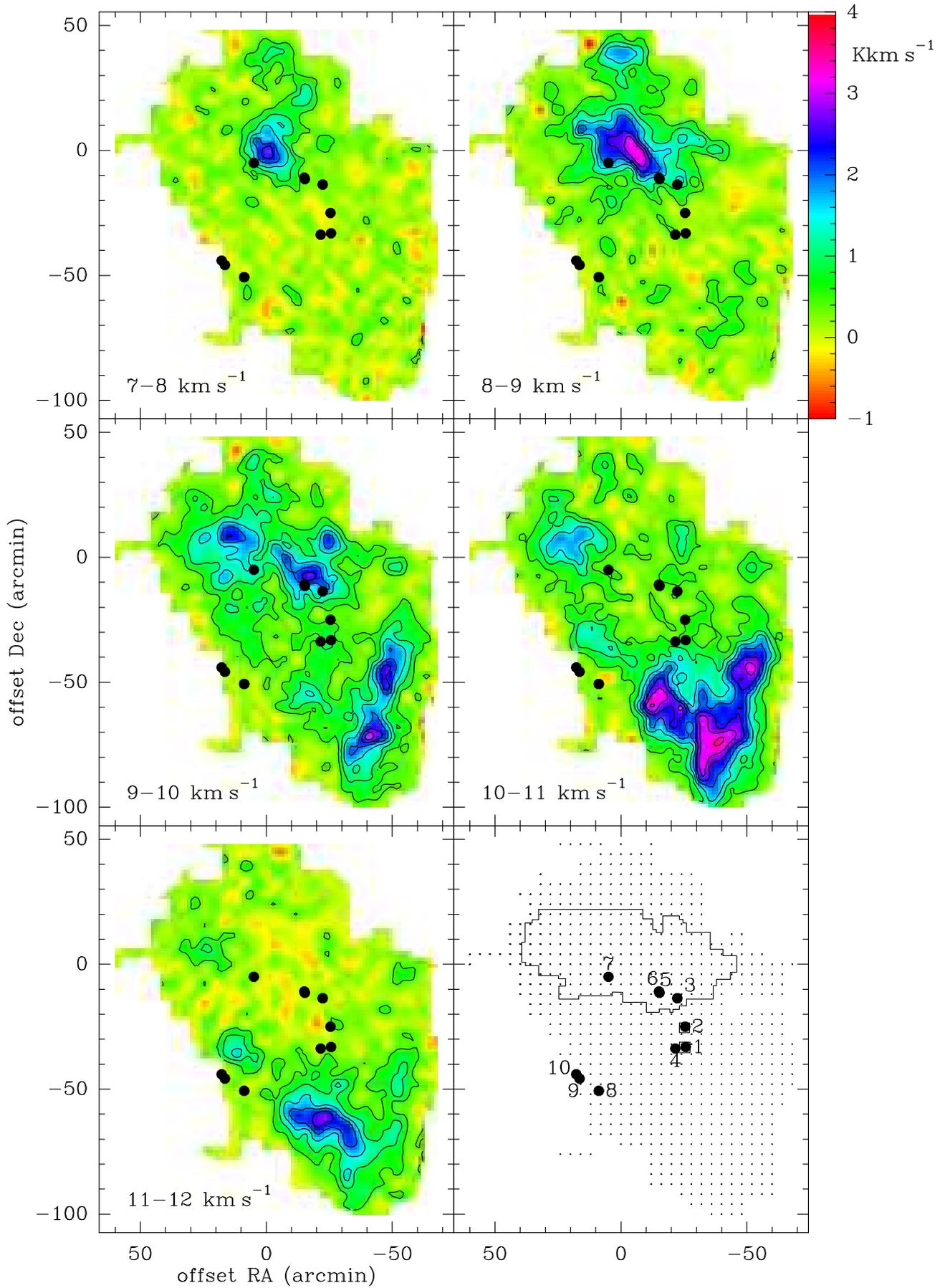}}
\caption{
KOSMA observations. 
$^{12}$CO(1--0) line emission \ITAstar, integrated in velocity 
intervals of 1~km\,s$^{-1}$ between 7 and 12~km\,s$^{-1}$. The lowest contour 
level and step are 0.5~Kkm\,s$^{-1}$. 
The observed positions are indicated in the
lower right panel. This panel also indicates the area observed with the SEST. 
The filled circles indicate the positions of the candidate emission-line 
stars, which are labeled with their numbers as in Table~\ref{targets}. 
Offsets as in Fig.~\ref{kosma_tp_ti}.}
\label{kosma_channel}
\end{figure*}

Following the criteria for the classification of late-type stars listed by
Turnshek et al. (\cite{turnshek}) and by comparison with the examples 
shown therein, we were able to roughly classify six of the stars as type 
M1~V - M4~V.
Star Ha4 (see the full spectrum in Fig.~\ref{fullspectra4} in the appendix; 
on line version only) 
is clearly not 
a late-type dwarf star. Through visual comparison with spectra 
from the atlas of Jacoby et al. (\cite{jacoby}) and with spectra from the 
SAO-FAST spectrograph\footnote{http://tdc-www.harvard.edu/cgi-bin/arc/fsearch} 
we determine this star to be of type F7-G1~V.

Possibly Ha4 should not have been included in the sample. Looking 
at the grism-spectrum from the Astroscan measurement (Sect.~\ref{red3p6}), 
this star appears to have a much bluer spectrum than 
the others, confirming our {\it DOLORES} spectrum. The feature that we 
originally identified as H$\alpha$ emission could have been a noise spike or 
be caused by dust on the grism-plate.

\subsection{KOSMA - CO \label{kosma}}

All molecular line spectra were reduced using the CLASS reduction software, 
developed by the Observatoire de Grenoble and IRAM. 
It was necessary
to subtract a higher order polynomial from some of the frequency-switched 
spectra, but because only narrow lines are present in the cloud, this did not
affect the emission profiles.

Figure~\ref{kosma_tp_ti} shows the distribution of the $^{12}$CO(1--0) emission
of the whole cloud, observed with the KOSMA 3-m telescope. One can 
distinguish two main peaks 
in the northeast and southwest parts of the cloud, respectively.
These peaks were already visible in the undersampled $^{12}$CO(1--0) map from 
Magnani et al. (\cite{mbm}), and are seen to break up into a 
number of smaller clumps.
The channel maps in 
Fig.~\ref{kosma_channel} show that the northeastern peak has a velocity 
of about 8.5~km\,s$^{-1}$, whereas the southwestern peak is at 10.5~km\,s$^{-1}$.
However, both peaks show internal velocity structure. The line width in the
area with the strongest emission is about 1.8~km\,s$^{-1}$.
The total emission in the region observed with the KOSMA telescope (using 
\ITmb) is 33999.2~Kkm\,s$^{-1}$arcmin$^2$, implying a CO-luminosity 
$L_{\rm CO} = 41.4$~Kkm\,s$^{-1}$pc$^2$ for a distance of 120~pc. 
Using the empirical relation $X = N(H_2)/\int T_{\rm mb}(CO) {\rm dv} = 
1.8 \times 10^{20}$~cm$^{-2}$(Kkm\,s$^{-1}$)$^{-1}$ (Dame et al.~\cite{dame}), 
we find a mass $M_{\rm CO} = 3.9 L_{\rm CO}$ = 162~\Msol, which includes  
a factor of 1.36 to account for helium.

\begin{figure}
\resizebox{9cm}{!}{\rotatebox{270}{\includegraphics{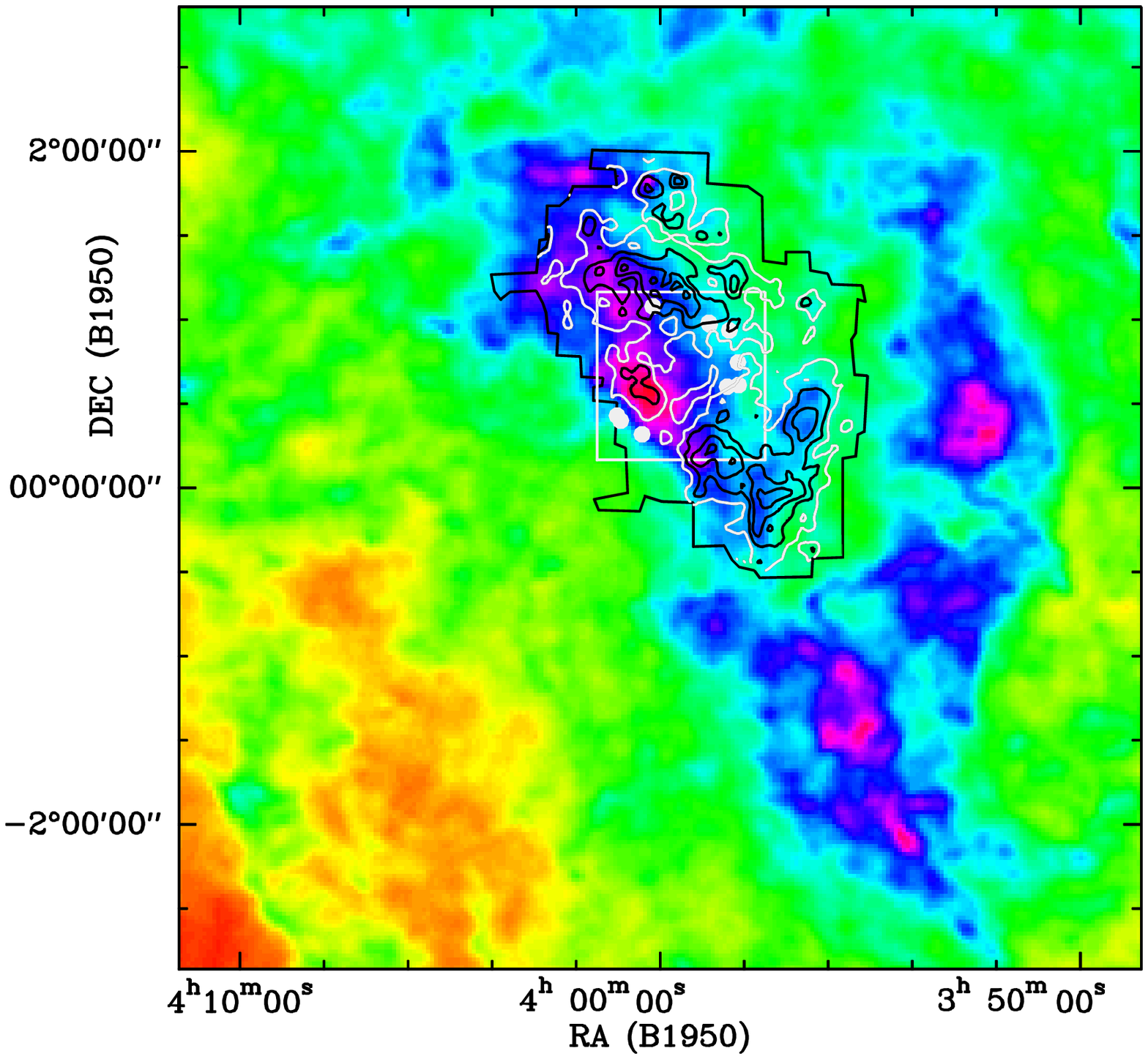}}}
\caption{Map of the IRAS 100~$\mu$m emission (data from NASA/JPL-Caltech). 
Contours: integrated 
$^{12}$CO(1--0) emission (\ITAstar; KOSMA observations; lowest contour and step 
2~K\kms; lowest contour drawn in white). The black irregular outline indicates
the mapped region. The white square shows the location and extent of the grism 
field, and 
the white filled circles indicate the candidate H$\alpha$ emission-line stars.}
\label{iras-grism-ha-kosma}
\end{figure}

The data shown here represent the first fully sampled map of this translucent 
cloud, whose basic morphology was first delineated by Magnani et al. 
(\cite{mbm}).  
However, a perusal of this particular cloud using the Schlegel et al. 
(\cite{schlegel}) dust map (Fig.~\ref{dustmap}) shows an extension 
to the west and south of the cloud. This is seen more clearly in 
Fig.~\ref{iras-grism-ha-kosma}, which shows the 
IRAS 100~$\mu$m map (or rather the improved reprocessed version of it: 
``IRIS''; from JPL-IPAC\footnote{http://irsa.ipac.caltech.edu/data/IRIS}), 
where we also 
indicate the KOSMA CO-contours, the location of the ESO 3.6-m grism-field 
(Sect.~\ref{obs3p6}), and the position of the H$\alpha$ candidate 
emission-line stars.
It is not known whether the 100~$\mu$m emission complexes to the west and 
south are also associated with CO emission. The southern high-latitude CO 
survey on a 1-degree sampling grid by Magnani et al. (\cite{magnani00}) shows 
CO emission from only two positions in this essentially unmapped region.
Although the dust emission from the southwestern clouds was not mapped in CO, 
as can be seen from Fig.~\ref{dustmap}, the dust column density (and, 
thus, the gas column density) is significantly lower than in MBM~18. 
In any case, our spectra show that in our map of MBM~18 we have reached the 
edges of that cloud, and an eventual molecular cloud associated with the 
other dust emission complexes will be separate (though likely related) entities.

\begin{figure*}
\resizebox{18cm}{!}{\includegraphics{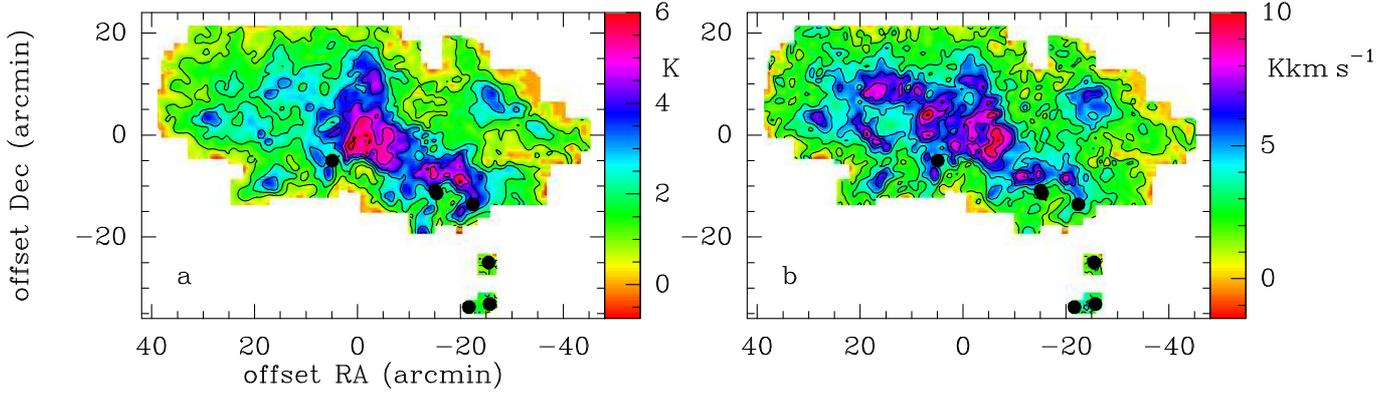}}
\caption{
{\bf a.}\ Distribution of peak $T_\mathrm{A}^*$ of $^{12}$CO(1--0) towards 
LDN~1569 obtained from SEST observations obtained from Gaussian fits to the 
spectra. Contour levels are 1 -- 6~K in steps of 1~K.
{\bf b.}\ as a., but for the line-integrated emission \ITAstar. 
Contour levels are 1.5 -- 10.5~Kkm\,s$^{-1}$ in steps of 1.5~Kkm\,s$^{-1}$. 
The filled circles indicate the positions of the candidate emission-line stars. 
Offsets as in Fig.~\ref{kosma_tp_ti}.
}
\label{sest_tp_ti}
\end{figure*}

\begin{figure*}
\resizebox{17cm}{!}{\includegraphics{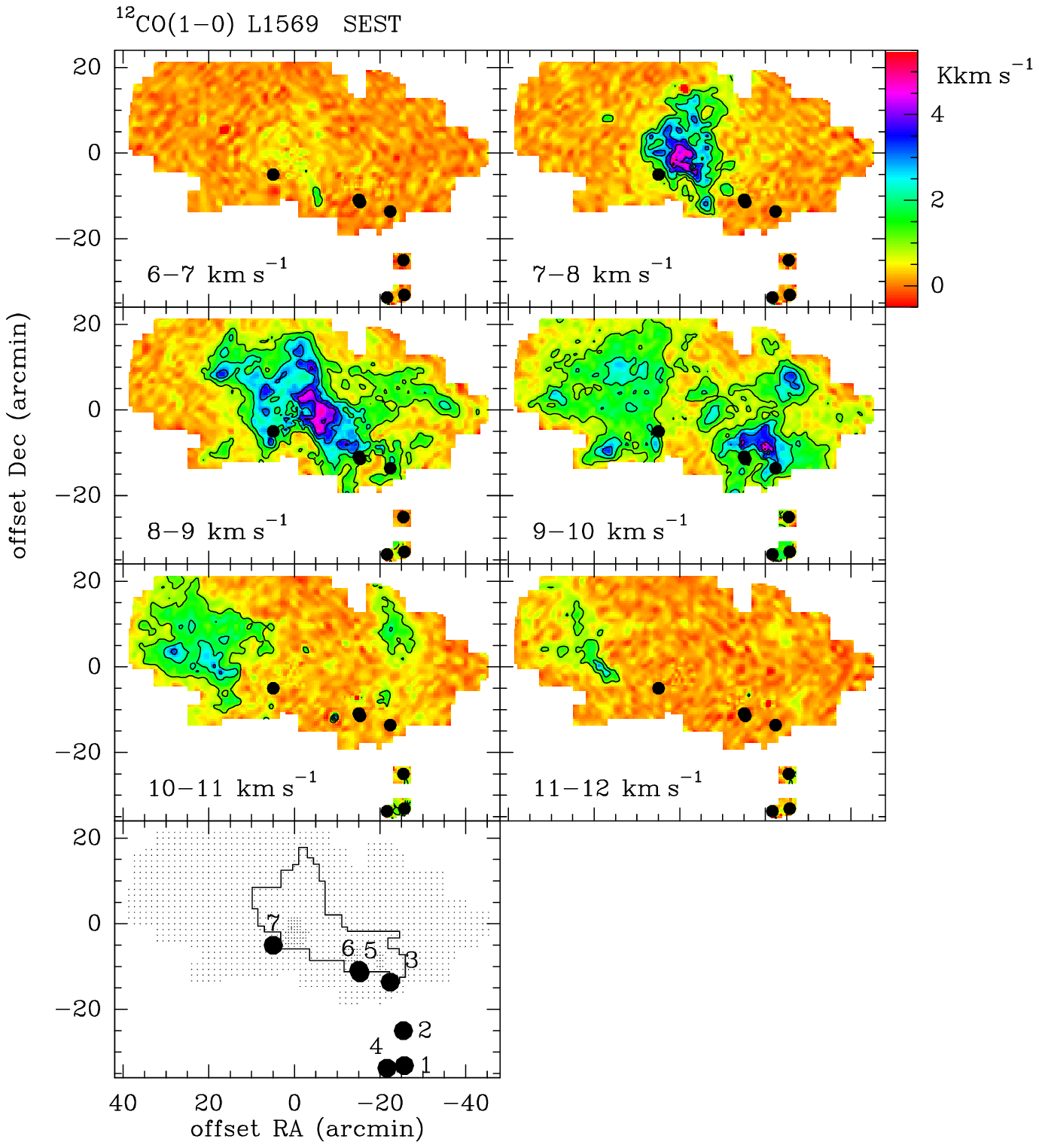}}
\caption{
SEST observations. 
$^{12}$CO(1--0) line emission \ITAstar\ integrated over velocity intervals 
of 1~km\,s$^{-1}$ between 6 and 12~km\,s$^{-1}$. Lowest contour level and step 
are 1~Kkm\,s$^{-1}$. The observed positions are indicated in the
lower panel. This panel also indicates the area observed with the SEST in 
$^{13}$CO(1--0). The filled circles indicate the positions of 
the candidate emission-line stars, which are labelled with their numbers as 
in Table~\ref{targets}.
Offsets as in Fig.~\ref{kosma_tp_ti}.}
\label{sest_channel}
\end{figure*}

\subsection{SEST - CO\label{sest}}

Figure~\ref{sest_tp_ti} shows the distribution of the $^{12}$CO(1--0) emission
of the northeastern peak, observed with the SEST. Also here we observe 
a high degree of clumpiness. The angular 
resolution of the SEST at 115~GHz corresponds to 0.027~pc at a 
distance of 120~pc.
Figure~\ref{sest_channel} shows that the central area has 
a velocity of
about 8~km\,s$^{-1}$, whereas the eastern and western parts are at a higher
velocity. The line width in the area of the $^{12}$CO peaks is 1.2~km\,s$^{-1}$.
The total emission in the region observed with the SEST (using \ITmb) is 
11808.3~Kkm\,s$^{-1}$arcmin$^2$, implying $L_{\rm CO} = 14.4$~Kkm\,s$^{-1}$pc$^2$ 
for a distance of 120~pc. This corresponds to a mass of 56~M$_\odot$ 
($X=1.8 \times 10^{20}$~cm$^{-2}$(Kkm\,s$^{-1}$)$^{-1}$, and including a factor 
of 1.36 to account for helium).\hfill\break\noindent
The ratio of $T_\mathrm{mb}$ $^{12}$CO(2--1) to (1--0) is about 1.0 (however, 
too few positions are observed in (2--1) to convolve these data to
the (1--0) angular resolution).

\begin{figure*}
\resizebox{18cm}{!}{\includegraphics{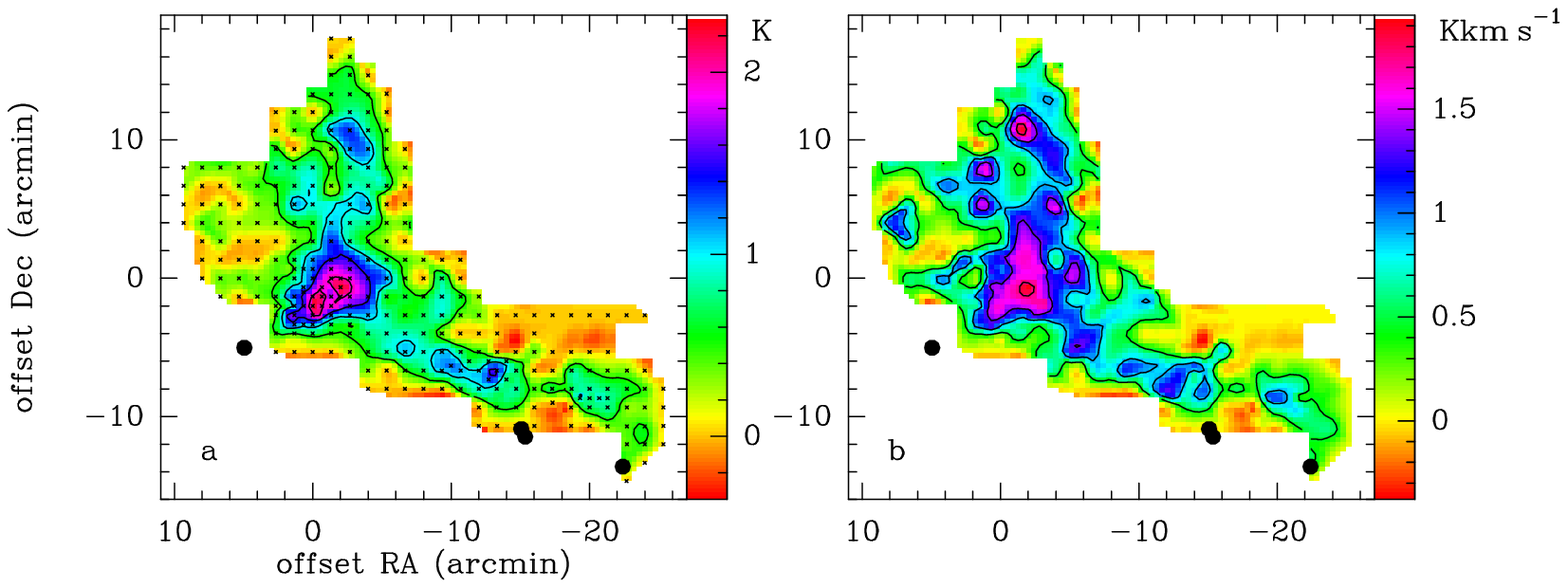}}
\caption{
{\bf a.}\ Distribution of peak $T_\mathrm{A}^*$ of $^{13}$CO(1--0) towards 
LDN~1569 obtained from SEST observations and Gaussian fits to the 
spectra. Contour levels are 0.5 to 2.0~K in steps of 0.5~K. The crosses 
indicate the observed positions.
{\bf b.}\ as a., but for the line integrated emission \ITAstar. Contour 
levels are 0.45 to 1.9~Kkm\,s$^{-1}$ in steps of 0.45~Kkm\,s$^{-1}$. The filled 
circles indicate the positions of the candidate emission-line stars. 
Offsets as in Fig.~\ref{kosma_tp_ti}.
}
\label{sest13co_tp_ti}
\end{figure*}

Figure~\ref{sest13co_tp_ti} shows the distribution of the $^{13}$CO(1--0) 
emission of the northeastern peak, observed with the SEST. 
This region contains one main peak and several smaller peaks, some of 
which are located along a ridge.
The line width at the positions with the strongest emission is 
0.75~km\,s$^{-1}$.
The $^{13}$CO(1--0) emission is much less extended than the $^{12}$CO(1--0) 
emission (cf. Fig.~\ref{sest_tp_ti}), a consequence of the low opacity of 
MBM~18 and clouds like it. Assuming LTE and \Tex = 10~K, we find a mass of 
5~\Msol\ from the $^{13}$CO observations (or $\sim 10$\% of the mass derived 
from the $^{12}$CO data), and $^{13}$CO optical depths of typically a few 
tenths, 
with a maximum of 0.6 in the core near offset 0,0 (see Fig.~\ref{nh3_13co}).
The $^{13}$CO column density at the location of peak optical depth is about 
$3.2 \times 10^{15}$~cm$^{-2}$. If we apply the standard relation 
$N$(H$_2$) = $5 \times 10^5 N$($^{13}$CO) (Dickman \& Clemens~\cite{dickman}) 
also to translucent clouds, the column density of H$_2$ $\approx 1.6 \times 
10^{21}$~cm$^{-2}$. Because $N$(H$_2$) = $10^{21} \times A_{\rm V}$ (Bohlin et 
al.~\cite{bohlin}, and assuming the gas-to-dust ratio in MBM~18 is the same 
as for the local interstellar medium), this implies a visual extinction 
$A_{\rm V}$ of about 1.6 mag at the peak of the $^{13}$CO core. 
The visual extinction can be derived independently of the 
Schlegel et al. (\cite{schlegel}) colour excess data (cf. Fig.~\ref{dustmap}) 
and the standard value of the total-to-selective absorption ratio $R_{\rm V}$ = 
$E$(B--V)/$A_{\rm V}$ of 3.1. At the position of the $^{13}$CO core, over a 
circular region with a diameter of 6\pam1, we find $A_{\rm V} \approx 1.8$~mag.

\begin{table}
\caption{Line parameters.}
\label{linepar}
\begin{flushleft}
\begin{tabular}{crrcccc}
\hline
\noalign{\smallskip}
line & \multicolumn{2}{c}{offset} & $T_{\rm mb}$ & $V_{\rm lsr}$ & $\Delta V$ & $\int T_{\rm mb} {\rm dv}$  \\
 &  RA & Dec & & & & \\ 
 & \multicolumn{2}{c}{arcmin} & K & km\,s$^{-1}$ & km\,s$^{-1}$ & Kkm\,s$^{-1}$ \\
\noalign{\smallskip}
\hline
\noalign{\smallskip}
CO(1--0)$^a$        & $-$2.7 & 0.7 & 7.35 & 7.93 & 1.38 & 10.79 \\
$^{13}$CO(1--0)$^a$ & $-$2.7 & 0.7 & 2.40 & 7.79 & 0.91 & 2.31   \\
CO(2--1)$^a$        & $-$1.3 & $-$1.3 & 4.94 & 8.00 & 2.18 & 11.50 \\
$^{13}$CO(2--1)$^a$ & $-$1.3 & $-$1.3 & 1.44 & 7.90 & 1.15 & 1.77 \\
C$^{18}$O(1--0)$^a$ & $-$1.3 & $-$1.3 & 0.31 & 7.72 & 0.39 & 0.13  \\
CS(2--1)$^a$        & $-$1.3 & $-$1.3 & ...$^{\dagger}$ & ...  & ... & ... \\
$^{13}$CO(2--1)$^a$ & 0.0 & 0.0 & 1.31 & 7.96 & 1.13 & 1.57 \\
$^{13}$CO(2--1)$^a$ & $-$13.3 & $-$6.7 & 0.67 & 9.24 & 0.89 & 0.63 \\
CO(2--1)$^b$        & 0.0 & 0.0 & 4.74 & 7.88 & 1.16 & 5.87 \\
CO(2--1)$^b$        & 9.37 & $-$32.93 & 0.72 & 11.20 & 1.53 & 1.17 \\
CO(2--1)$^b$        & 5.98 & $-$36.30 & 1.44 & 10.68 & 1.08 & 1.66 \\
CO(2--1)$^b$        & 2.58 & $-$39.67 & 1.10 & 10.88 & 1.71 & 1.99 \\
NH$_3$(1,1)$^c$     & $-$2.3 & 0.5 & 0.20 & 7.88 & 0.91 & 0.19 \\
NH$_3$(2,2)$^c$     & $-$2.3 & 0.5 & 0.05 & 7.82 & 0.27 & 0.02   \\
\noalign{\smallskip}
\hline
\end{tabular}
\smallskip\noindent
\\
$^a$ SEST data; $^b$ JCMT data; $^c$ Effelsberg data, average spectra of the 
strongest 3 positions; $^{\dagger}$ rms = 0.064~K\\
\end{flushleft}
\end{table}

\subsection{Effelsberg - NH$_3$}

\begin{figure}
\resizebox{9cm}{!}{\rotatebox{0}{\includegraphics{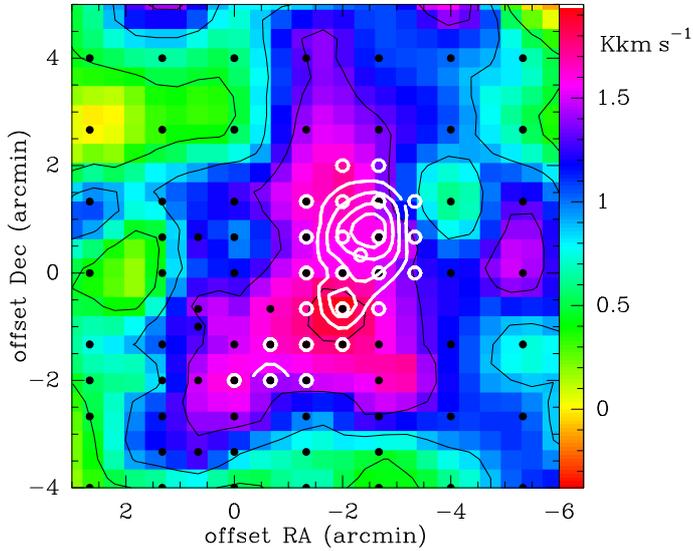}}}
\caption{
{\bf a.}\ Distribution of integrated NH$_3$(1,1) emission (white contours) 
towards LDN~1569  (Effelsberg data) superposed on the integrated 
$^{13}$CO(1--0) 
emission (colour plot; black contours) observed with the SEST.
The NH$_3$ contour levels are 0.05 (0.05) 0.20~Kkm\,s$^{-1}$ ($\int 
T_\mathrm{mb}$dv) 
and the $^{13}$CO \ITAstar\ contour levels 0.45 (0.45) 1.8~Kkm\,s$^{-1}$.
Black dots indicate positions observed in $^{13}$CO and white circles those
observed in NH$_3$.
}
\label{nh3_13co}
\end{figure}

The NH$_3$(1,1) line was observed and detected near the core of the cloud 
area that was mapped in CO with
the SEST. The NH$_3$ emission is plotted superimposed on the $^{13}$CO(1--0) 
emission in Fig.~\ref{nh3_13co}. The strongest NH$_3$ emission is slightly 
displaced from the $^{13}$CO peak. Line parameters of the
average spectrum of the three strongest positions are given in 
Table~\ref{linepar}, together with the parameters of other lines observed at
that and other positions. A fit to the hyperfine components indicates that the 
optical depth of the line is small ($<$0.1). The line width corrected for 
hyperfine components is 0.64$\pm$0.08~km\,s$^{-1}$.
The fact that NH$_3$ has been detected suggests that fairly dense
material is present, which means that at least some of the necessary 
conditions for star formation are present 
in this cloud. (Note that CS was not detected, however.)  
The rotation temperature, derived from the NH$_3$(1,1) and (2,2) results, 
following Harju et al. (\cite{harju}) is 39$\pm$10~K, and the NH$_3$ column 
density $(1.4 \pm 0.4) \times 10^{13}$~cm$^{-2}$. 

\subsection{Medicina - H$_2$O masers}

\begin{figure}
\resizebox{9cm}{!}{\includegraphics{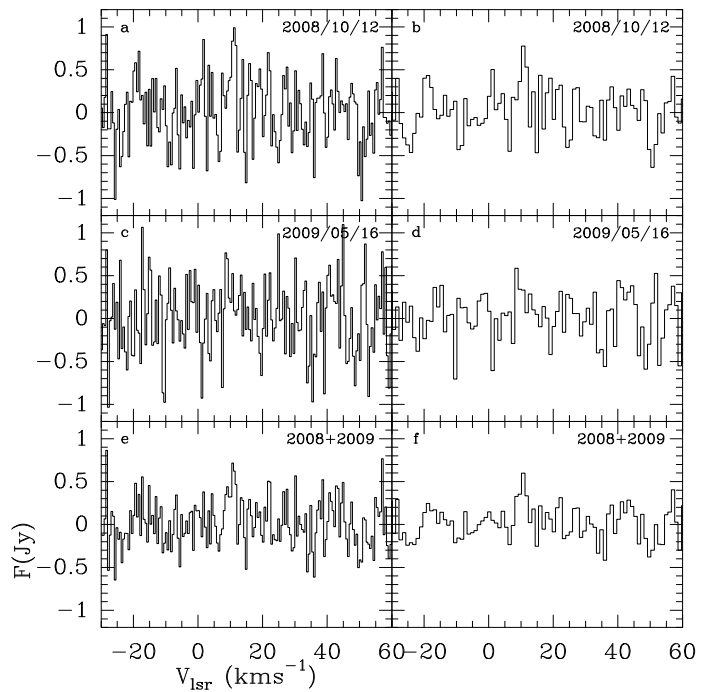}}
\caption{
{\bf a.}\ October 2008 spectrum at 22~GHz of candidate H$\alpha$ 
emission-line star Ha7. The spectrum is smoothed to a resolution of 0.53~\kms.
{\bf b.}\ As a, but smoothed to a resolution of 1.05~\kms. 
{\bf c.}\ As a, but for the May 2009 spectrum.
{\bf d.}\ As b, but for the May 2009 spectrum.
{\bf e.}\ Average spectrum at 22~GHz of the 2008 and 2009 observations of 
Ha7, smoothed to a resolution of 0.53~\kms. 
{\bf f.}\ As e, but smoothed to a resolution of 1.05~\kms. 
}
\label{star5-h2o}
\end{figure}

All data were reduced with the CLASS-package. 
Polynomial baselines were removed and spectra were 
averaged for each position. The results are shown in Table~\ref{masers}. 
The spectra taken towards Ha7 in Oct. 2008 and May 2009 showed a possible 
detection when smoothed to a lower resolution (see Fig.~\ref{star5-h2o}); 
results of a Gaussian-fit to the average spectra are reported in the last 
two lines of Table~\ref{masers}. 
To confirm the reality of this feature we reobserved this position in January 
2011 for 45~minutes on-source. No signal was detected; averaging {\it all} 
spectra pertaining to Ha7, including those from 2011, shows no detection. 
Water masers are notoriously variable, especially those associated with 
lower-mass stars (Felli et al.~\cite{felli} and references therein) and this 
non-detection more than 1.5~years after the original observation does not 
imply that the tentative detection of maser emission in the 2008/2009 
observations is not real.

\begin{table}
\caption{Medicina H$_2$O maser observations.}
\label{masers}
\begin{flushleft}
\begin{tabular}{lccccc}
\hline
\noalign{\smallskip}
\multicolumn{1}{c}{Position} & \multicolumn{1}{c}{date} 
 & \multicolumn{1}{c}{rms$^1$} 
 & \multicolumn{1}{c}{$F$}
 & \multicolumn{1}{c}{$\Delta V$} 
 & \multicolumn{1}{c}{$V_{\rm lsr}$} \\
\multicolumn{1}{c}{} & \multicolumn{1}{c}{(y-m-d)} 
 & \multicolumn{1}{c}{(Jy)} 
 & \multicolumn{1}{c}{(Jy)}
 & \multicolumn{1}{c}{(\kms)}
 & \multicolumn{1}{c}{(\kms)} \\
\noalign{\smallskip}
\hline
\noalign{\smallskip}
Ha1 & 2008-10-11 & 1.14 & ... & ... & ...  \\
      &2009-05-16 & 1.74 & ... & ... & ...  \\
      &2011-02-20 & 1.55 & ... & ... & ...  \\
Ha2 &2008-10-12 & 1.02 & ... & ... & ...  \\
      &2009-05-16 & 1.08 & ... & ... & ...  \\
      &2011-02-20 & 1.69 & ... & ... & ...  \\
      &2011-03-18 & 1.13 & ... & ... & ...  \\
Ha5 &2008-10-10 & 1.12 & ... & ... & ...  \\
      &2009-04-02 & 1.35 & ... & ... & ...  \\
      &2009-05-16 & 0.99 & ... & ... & ...  \\
      &2011-02-20 & 1.28 & ... & ... & ...  \\
      &2011-03-18 & 0.60 & ... & ... & ...  \\
Ha6 &2009-05-16 & 1.07 & ... & ... & ...  \\
      &2011-02-20 & 1.42 & ... & ... & ...  \\
Ha7 &2008-10-12 & 0.92 & ... & ... & ...  \\
      &2009-05-16 & 1.10 & ... & ... & ...  \\
      &2011-01-19 & 0.36 & ... & ... & ...  \\
      &2011-02-20 & 1.24 & ... & ... & ...  \\
      &2011-03-18 & 1.02 & ... & ... & ...  \\
Ha8 &2008-10-10 & 0.99 & ... & ... & ...  \\
      &2009-05-16 & 1.08 & ... & ... & ...  \\
      &2011-02-20 & 2.13 & ... & ... & ...  \\
Ha9 &2008-10-11 & 1.51 & ... & ... & ...  \\
      &2009-05-16 & 1.03 & ... & ... & ...  \\
Ha10 &2008-10-11 & 1.27 & ... & ... & ...  \\
      &2009-05-16 & 1.08 & ... & ... & ...  \\
\hline
\noalign{\smallskip}
Ha7$^2$&2008+09&0.28&0.55&$2.78\pm 0.90$&$10.46\pm 0.41$ \\
Ha7$^3$&2008+09&0.18&0.53&$2.88\pm 0.89$&$10.45\pm 0.42$ \\
\noalign{\smallskip}
\hline
\end{tabular}
\smallskip\noindent
\\
$^1$\ rms noise per channel of width 0.132~\kms.\\
$^2$\ Line parameters from a Gaussian fit to the average of the spectra from 
2008 and 2009, smoothed to a resolution of 0.53~\kms.\\ 
$^3$ As 2, but smoothed to a resolution of 1.05~\kms.
\\
\end{flushleft}
\end{table}

\begin{figure}
\resizebox{9cm}{!}{\includegraphics{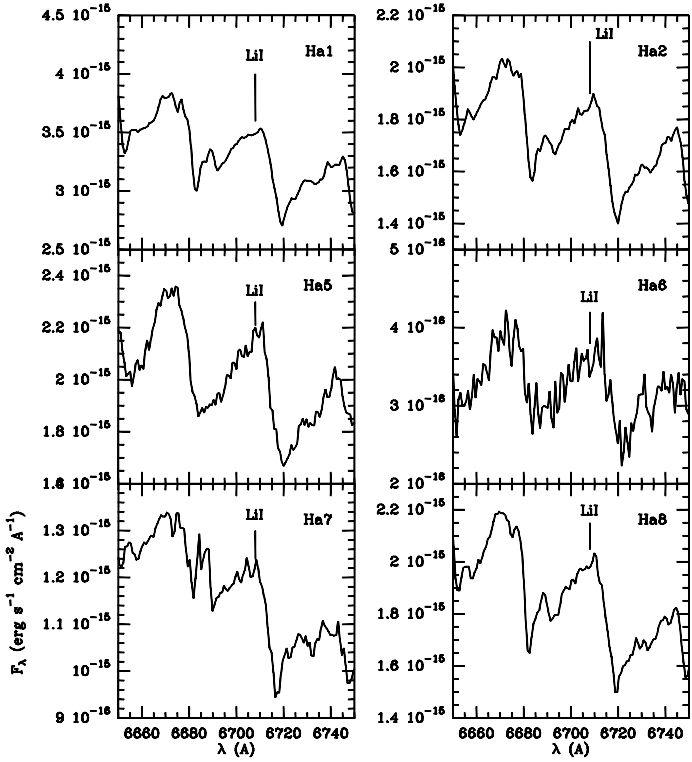}}
\caption{
Detail of the spectra of the targets identified in each panel, 
showing the wavelength region around LiI (as indicated). For all stars
except Ha1 and Ha7 (which have no H$\alpha$), the wavelength scale was 
adjusted by shifting the H$\alpha$ line to its rest frequency (see text). 
}
\label{LiIdetail}
\end{figure}

\begin{figure*}
\sidecaption
\includegraphics[width=12cm]{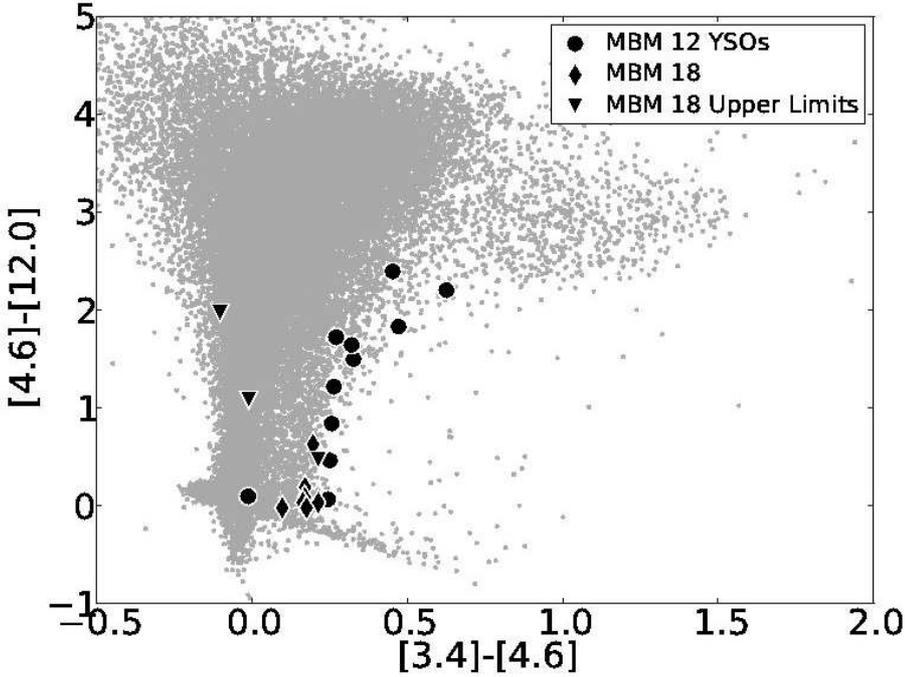}
\caption{
Colour-colour diagram based on WISE data. The grey dots are 
point sources seen projected against all clouds in the MBM catalogue. 
Most in the swarm have the colours of (spiral) galaxies, while stars and 
elliptical galaxies have colours near zero. The diamonds identify 
the candidate H$\alpha$ emission-line stars found in our grism field 
(Table~\ref{targets}), with upper limits indicated by upside-down 
triangles. 
For comparison, the circles indicate the YSOs 
found in MBM12 (Luhman~\cite{luhman}).}
\label{wise}
\end{figure*}

\begin{figure}
\resizebox{9cm}{!}{\includegraphics{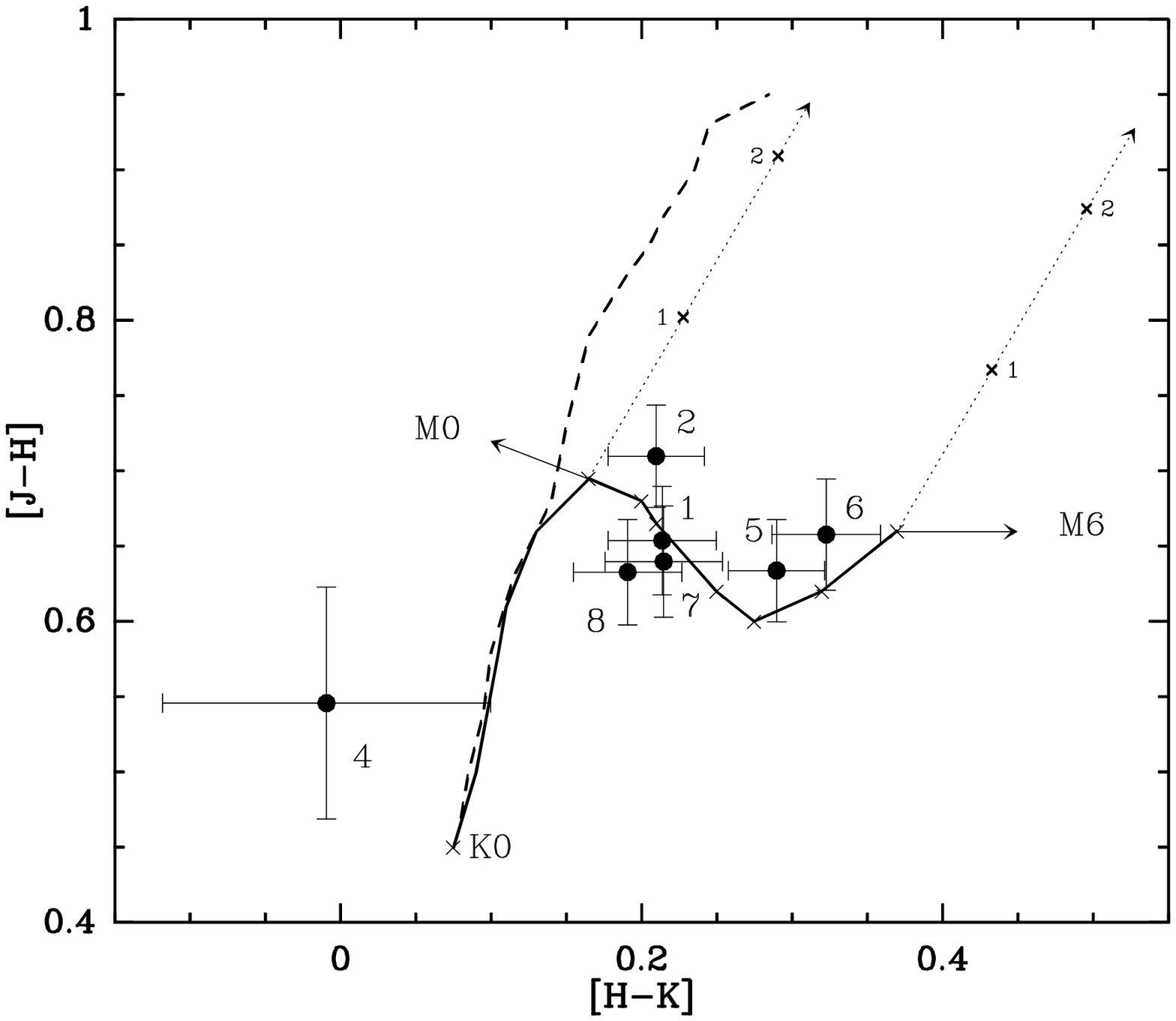}}
\caption{
Colour-colour diagram of the observed stars (2MASS colours transformed to 
the Bessel \& Brett (\cite{bessel} system; see text). The intrinsic 
main sequence (drawn; shown for spectral types K0 and later) and giant branch 
(dashed; from type G4 at the bottom to M5 at the top) are taken from Bessel 
\& Brett (\cite{bessel}). 
Along the main sequence we indicate the locations of unreddened stars 
of type M0~V to M6~V with crosses and label two of these. 
The direction of interstellar reddening is indicated by the dotted lines 
starting at types M0 and M6, where black crosses indicate 
values of visual extinction $A_{\rm V} =1$ and 2~mag.
}
\label{col-col}
\end{figure}

\section{Discussion \label{discus}}

\subsection{Stars \label{stars}}

We derived more accurate spectral types (for theM-stars) from our spectra 
using the relation
S$_p = -10.775$ (TiO5) + 8.200 (Cruz \& Reid~\cite{cruz}; see their Fig.~3 and 
Table~4), with TiO5 a spectral index 
describing the strength of the TiO band at $7128-7135$~\AA, compared with 
the continuum at $7042-7046$~\AA. 
The uncertainty of the spectral types derived using this relation is 0.5 
subtype. The results are presented in Table~\ref{tngresults}.
The other columns in this table contain the following information: 
the star name (Col.~1); an indication of whether H$\alpha$ emission is 
(Y) or is not (N) present (Col.~2); if ``Y'', Col.~3 gives the equivalent 
width of this line. Col.~4 gives an indication (Y/N) of the presence of the 
LiI 6708~\AA\ absorption line. In Cols.~5-10 we give the K-magnitude, the 
(H$-$K) 
and (J$-$H) colours, and their uncertainties, respectively. Magnitude and 
colours are originally from the 2MASS survey (see Table~\ref{targets}), and 
were transformed to the Bessel \& Brett (\cite{bessel}) system using 
Eqs. (A1)--(A4) in Carpenter (\cite{carpenter}). 
Finally, Cols.~11, 12, and 13 list 
the spectral types, the extinction in the K-band ($A_{\rm K}$), and the 
distance of the stars, respectively. Several of these parameters will be 
determined in the remainder of this section.

As described in Sect.~3.6, the wavelength calibration of the 
spectra was corrected using measurements of sky lines. 
After this, the H$\alpha$ lines in the spectra do not always lie exactly at 
the rest frequency. Any remaining difference  
between observed and laboratory wavelengths is due to the radial velocities 
of the stars with respect to the telescope at the time of observation. 
For the H$\alpha$ line this difference ranges from 
$-1.3$\AA\ (Ha8) to +1.1\AA\ (Ha6); Ha1 and Ha7 have no clearly visible 
H$\alpha$ emission or absorption, and an offset could not be determined.

Clear H$\alpha$ emission was 
detected in all stars except Ha1 and Ha7. The wavelength region around 
LiI is shown in more detail in Fig.~\ref{LiIdetail}. Here we have adjusted 
the wavelength scale by forcing the observed H$\alpha$ line to 
its rest frequency (except for Ha1 and Ha7, the spectra of which have no 
H$\alpha$ line). None of the stars show strong 
LiI-absorption; there are perhaps hints of the line in Ha5, Ha7 and Ha6, 
although the spectrum of latter star is rather noisy.

Optical spectra of T Tauri stars show a stellar absorption line spectrum, 
resembling that of a late-type (K or M) dwarf or subgiant photosphere, 
superimposed by an emission line spectrum (Appenzeller \& 
Mundt~\cite{appenzeller}). Furthermore, a 
distinguishing characteristic of all T Tauri stars is strongly enhanced LiI 
absorption (weak or absent in most evolved late-type stars; Appenzeller \& 
Mundt~\cite{appenzeller}, and references therein). It is clear that the 
stars observed by us are not T Tauri stars, but M-type dwarf stars.

Both M-dwarfs and T Tauri stars (especially weak-lined ones, i.e. H$\alpha$ 
absolute equivalent width 
less than $\sim 10$\AA) often show X-ray emission (e.g. Neuh\"auser et 
al.~\cite{neuhauser}). The ROSAT All-Sky Survey data for MBM~18 were 
analysed by Hearty (\cite{hearty}) following the procedure discussed by 
Neuh\"auser et al.~\cite{neuhauser}). For two of our stars Hearty found an 
X-ray 
counterpart: Ha5 (RXJ0401.4+0106) and Ha8 (RXJ0403.0+0027a). Only the latter 
appears also in the 
Second ROSAT PSPC Catalog (ROSAT 2000; 2RXP 
J040300.2+002748) and in the WGA-catalogue (White et al.~\cite{white}; 1WGA 
J043.0+0027). The X-ray counterpart of Ha5 is apparently just below the 
detection threshold established by these catalogues. Hearty (\cite{hearty}) 
obtained an optical spectrum of Ha5 and classified it as a type M3, similar 
to what we find (Table~\ref{tngresults}). 

The non-T Tauri nature of our target stars is confirmed by comparing their 
location with those of the young stellar objects (YSOs) in MBM~12, in a 
colour-colour diagram constructed using WISE 
observations, shown in Fig.~\ref{wise}. The WISE mission 
(Wright et al.~\cite{wright}) has made a sensitive all-sky survey in four 
mid-IR bands. In a colour-colour plot as shown in Fig.~\ref{wise} (cf. Fig.~12 
in Wright et al.~\cite{wright}), stars (and early-type galaxies) have colours 
near zero, spirals are red in the [4.6]--[12.0] colour, while other 
extra-galactic objects such as Seyfert galaxies and active galactic nuclei 
are red in both 
colours. In Fig.~\ref{wise} we plot the colours of point sources seen projected 
against all clouds in the MBM-catalogue as grey dots. These point sources 
fall into two broad categories: stars and galaxies; because of the high 
(absolute) galactic latitude of the MBM-clouds and their low extinction, it 
is not a surprise that we find so many extra-galactic objects at those 
locations. The H$\alpha$-stars from Table~\ref{targets} are plotted as 
diamonds. All lie in the colour range of normal stars, except perhaps for Ha4, 
which however has a lower limit to the magnitude in the 12~$\mu$m band, and 
thus an upper limit to the [4.6]--[12.0] colour (we will further remark on 
this star below), as do Ha3 and Ha10 (neither of them was observed 
spectroscopically by us).
As a comparison we also show the locations of the certified YSOs in MBM~12 
(Luhman~\cite{luhman}; circles), which have a very different distribution 
from the stars we 
observed in MBM~18. This confirms that the MBM~18 stars do not have YSO 
characteristics, as is indeed borne out by their optical spectra.

In Fig.~\ref{col-col} we plot the locations of the target stars in the 
infrared colour-colour diagram. The main sequence and the giant branch are 
taken from Bessel \& Brett (\cite{bessel}). As in Fig.~\ref{wise}, the 
location of Ha4 is peculiar. Its spectrum is that of a normal main-sequence 
star, and as such it should lie between the two dotted lines that 
indicate the normal reddening band. We can exclude that Ha4 is a galaxy,
because 
the offset of the H$\alpha$ line 
from its rest-wavelength (0.87\AA) indicates a velocity with respect to the 
Local Standard of Rest of about $-90$~\kms. It could be a single-line 
spectroscopic binary: the companion is too weak to show up in the spectrum 
of the primary, but does affect the colours.

From this plot we can also 
obtain an estimate of the spectral types of the stars. These are found to 
agree with those determined from the relative strength of the TiO band, 
except for Ha2 and Ha8, for which Fig.~\ref{col-col} suggests M1 and M2, 
respectively. In the following we use the spectral types as given in 
Table~\ref{tngresults}, although the results do not change significantly 
(not even for Ha2 and Ha8) if we use those derived from Fig.~\ref{col-col} 
instead. 

From the observed and intrinsic colours (Cox~\cite{cox}) we find the colour 
excess and thus the extinction, by averaging 
$A_{\rm V} = 15.87{\rm E}(H-K)$ and 
$A_{\rm V} = 9.35{\rm E}(J-H)$; furthermore $A_{\rm K}=0.112A_{\rm V}$ (Rieke \& 
Lebofsky~\cite{rieke}).
As can be seen from Fig.~\ref{col-col}, the M-type stars are all very close 
to the 
main-sequence locus and the extinction will therefore be very small, and 
in fact for several stars we found the two colour excesses very close to zero 
but of opposite sign. In these cases we assumed $A_{\rm V} = A_{\rm K} = 0$. 
The distance modulus for a star with observed magnitude $m_{\rm K}$ and 
absolute magnitude $M_{\rm K}$ is $m_{\rm K} - M_{\rm K} = 5{\rm log}d -5 + 
A_{\rm K}$, from which we can derive the distance $d$. The absolute 
$M_{\rm K}$ magnitudes for the various M-type stars were found from the 
intrinsic $(V-K)$-colour (Cox~\cite{cox}) 
and the absolute 
visual magnitude $M_{\rm V}$ (Schmidt-Kaler~\cite{schmidt}). 

The results are summarised in Table~\ref{tngresults}. The first thing we 
notice is the wide range in distances of the stars, from 60 to 250~pc, and 
the low extinction values. 
The distance to MBM~18 has been determined by several authors, using 
a variety of methods such as star counts (Magnani \& de 
Vries~\cite{magnani86}; Chaplin et al.~\cite{chaplin}), photometry 
(Franco~\cite{franco}; Penprase~\cite{penprase92}), and spectroscopy of 
interstellar absorption lines (Penprase et al.~\cite{penprase90}; 
Penprase~\cite{penprase93}). It should be noted that virtually none of the 
stars studied in the literature are seen projected against that part of 
the cloud that has been detected in CO, which is where our target stars are 
located (in projection). Most authors agree more or less on a distance of 
$120-150$~pc, but the cloud has ``a patchy and highly irregular structure'' 
(Penprase~\cite{penprase93}). From a plot of visual extinction as 
a function of distance, Penprase (\cite{penprase92}) concluded that the 
MBM~18/LDN~1569 complex extends along the line-of-sight from about 140~pc 
to 300~pc. Regarding this latter statement, we note that 
it is highly unlikely that, as a molecular entity, MBM~18/LDN~1569 is 
a few parsec wide and 160~pc long, with the long axis aligned almost 
exactly along the line-of-sight. The region containing MBM~18 and MBM~16 (see 
Fig.~\ref{dustmap}) lies south of the Taurus-Aurigae dark cloud complexes, 
which have several absorbing layers at different distances (e.g., 
Magnani~\cite{magnani88}). It is likely that this complex distribution of 
absorbing layers continues south of Taurus into the region containing 
MBM~18 and MBM~16. Those clouds are likely associated with just one 
of those absorbing layers. 

On the other hand, from a similar study as that by Penprase 
(\cite{penprase92}) a considerably lower value 
($80 \pm 20$~pc) was found by Chaplin et al. (\cite{chaplin}). 
Magnani \& de Vries (\cite{magnani86}) from a comparison of theoretical 
and observed Wolf diagrams found a best fit for a distance of 50~pc (and 
an upper limit of 175~pc). In view of the above, all stars in 
Table~\ref{tngresults} are apparently located 
at least in the 
extended complex that Penprase (\cite{penprase93}) refers to; but Ha1, Ha8, 
and perhaps also Ha2, lie at the distance that the bulk of the molecular 
material of MBM~18 is located at. The visual extinction 
of our targets stars has a maximum value of about 0.5~mag, which is a 
typical average value for the cloud (e.g., Penprase~\cite{penprase93}). 
Four of the stars have no measurable extinction, among which Ha7 
at 252~pc is the most distant one in our sample.

When comparing the photometry of our stars with the pre-main-sequence 
isochrones of Palla \& Stahler (\cite{palla99}) we obtain a good fit for 
values of distances, extinction, and masses in accordance with the values 
listed in Table~\ref{tngresults}. These fits show that the stars have ages 
between 7.5 and 15~Myr: young enough to fit on pre-main-sequence isochrones, 
but old enough for Li to have disappeared (in low-mass stars lithium is 
effectively destroyed by convective mixing on timescales of a few Myr). 
Considering the unbound state of the molecular gas (see Sect.~\ref{molgas}), 
these stars are too old to have been formed in MBM~18, as presently constituted.

We thus find at least four active late-type stars in our grism field, with
spectral types down to dM4. We can derive the space density of M-dwarfs of
type dM0$-$dM4 from the luminosity function of M-dwarfs, derived by Zheng et
al. (\cite{zheng}) from HST star counts at $\mid$ {\sl b} $\mid > 17$\degr. 
According to Joy \& Abt (\cite{joy}), 11\% of stars in that range of spectral 
types are dMe. In the volume
of space sampled by us in the 1 sq. degree grism field (a truncated cone
between 80 and 300~pc - which is roughly the extent of molecular gas along
the line-of-sight towards MBM~18), we therefore expect about 6dMe stars of type
dM0$-$dM4, which agrees with our observations.

\begin{table*}
\caption{Results from the TNG observations for the late-type stars}
\label{tngresults}
\begin{flushleft}
\begin{tabular}{clclcccccccrr}
\hline
\noalign{\smallskip}
Star & H$\alpha$ & EW(H$\alpha$)$^{\dagger}$ & LiI & K$^a$ & $\sigma_{\rm K}$ & (H-K)$^a$ & $\sigma_{\rm H-K}$ & (J-H)$^a$ & $\sigma_{\rm J-H}$ & Sp.T. & $A_{\rm K}$ 
& $d$ \\
  &   & \AA  &  &\multicolumn{2}{c}{mag}& \multicolumn{2}{c}{mag} &
\multicolumn{2}{c}{mag} &       & mag       & pc \\
\noalign{\smallskip}
\hline
\noalign{\smallskip}
Ha1 & N & ... & N & 11.304 &0.026 & 0.214 &0.036 & 0.654 &0.036 & dM2.5 & 0.0 & 128 \\
Ha2 & Y & $-1.5$ & N & 11.076 &0.023 & 0.210 &0.032 & 0.710 &0.034 & dM3.5e & 0.0 & 109 \\
Ha5 & Y & $-4.0$ & Y? & 10.044 &0.021 & 0.290 &0.032 & 0.634 &0.034 & dM4.0e & 0.02 & 63 \\
Ha6 & Y & $-5.1$ & Y? & 10.988 &0.023 & 0.323 &0.036 & 0.658 &0.037 & dM4.0e & 0.06 &95 \\
Ha7 & N & ... & Y? & 12.580 &0.030 & 0.215 &0.039 & 0.640 &0.037 & dM1.5 & 0.0 & 252 \\
Ha8 & Y & $-3.0$ & N & 11.390 &0.026 & 0.191 &0.036 & 0.633 &0.035 & dM3.5e & 0.0 & 128 \\
\noalign{\smallskip}
\hline
\end{tabular}
\smallskip\noindent
\\
$^{\dagger}$\ By convention negative equivalent widths indicate emission lines\\
$^a$\ Magnitudes and colours transformed from 2MASS to the Bessel \& Brett 
(\cite{bessel}) system using Carpenter (\cite{carpenter}) \\
\end{flushleft}
\end{table*}

\subsection{Molecular gas \label{molgas}}

Using the KOSMA $^{12}$CO data, we determined the virial state of the cloud.
The number of detections (499) multiplied by the area of a map-element 
($4\times 4$~arcmin$^2$) can be used as an estimate of the cloud area. 
The 'effective radius' of the cloud is then $r_{\rm e} = 
\sqrt({\rm area}/\pi)$; for a representative distance of 120~pc we find 
$r_{\rm e} = 1.8$~pc. 
Gaussian profiles were fitted to the 499 detections and the variance of the 
velocity centroids was 0.88~\kms, equivalent to a $\Delta V$(FWHM) of 2.1~\kms. 
For a density distribution $\rho \propto r^{-2}$, 
$M_{\rm vir}({\rm M}_{\odot}) = 126 r_{\rm e} \Delta V^2$, where $r_{\rm e}$
is the radius of the cloud in pc and $\Delta V$ (\kms) is a measure of the 
velocity dispersion along the given line-of-sight.
We thus obtain a virial mass for the cloud $M_{\rm vir} \approx 1000$~\Msol.

In Sect.~\ref{kosma} we obtained a mass (from the integrated emission) for 
the cloud of 162~\Msol\ for a distance 
of 120~pc. It is therefore clear that MBM~18, like most high-latitude clouds, is
not gravitationally bound and should be breaking up on the sound crossing time
scale ($\sim$ 10$^6$~years). However, this does not mean that the cloud is
incapable of star formation. Gravitationally unbound clouds can 
contain clumps that are dense enough to be bound, and in which star formation 
might occur. The well-studied high-latitude cloud, MBM~12,
is not gravitationally bound either (Pound et al.~\cite{pound}), yet it is a
prolific star-forming region with at least a dozen young stellar objects
associated with the cloud (Luhman~\cite{luhman}). 

\smallskip
We ran CLUMPFIND\footnote{http://www.ifa.hawaii.edu/users/jpw/clumpfind.shtml} 
(Williams et al.~\cite{williams}) on the KOSMA observations, 
using 0.6~K for the lowest contouring value and the increment 
($\sim 2.9$ times the median rms in the data - see Sect.~2.3.1), 
and found 12 clumps, 
with parameters as listed in Table~\ref{kosmaclumps}. The combined mass 
in all clumps, 
as determined from their combined CO-luminosity, is $\sim 116$~\Msol, 
which is $\sim 72$\% of the total mass, determined from all emission (see 
Sect.~\ref{kosma}).

\begin{table*}
\caption{Parameters of the clumps identified in the KOSMA CO(1--0) data}
\label{kosmaclumps}
\begin{flushleft}
\begin{tabular}{rrrcrcrrcccrr}
\hline
\noalign{\smallskip}
Clump & $\Delta\alpha$ & $\Delta\delta$ &$T_{\rm mb}$ & 
\multicolumn{1}{c}{$V_{\rm lsr}$} & $\Delta V$
 & \multicolumn{1}{c}{$r_1$} & \multicolumn{1}{c}{$r_2$} & 
$r_1$ & $r_2$ & $L_{\rm CO}$ & $M_{\rm vir}$($r_1$) & $M_{\rm vir}$($r_2$) \\  
      & \multicolumn{2}{c}{(arcmin)} & (K) & \multicolumn{2}{c}{(\kms)} & 
\multicolumn{2}{c}{(arcmin)} & \multicolumn{2}{c}{(pc)} & K\kms pc$^2$ & 
\multicolumn{2}{c}{(\Msol)} \\
\noalign{\smallskip}
\hline
\noalign{\smallskip}
  1& $-$41.7& $-$73.8&  5.23& 10.62&  1.81&  9.19& 18.18& 0.321& 0.634& 5.726& 131.9& 260.9 \\
  2& $-$14.6& $-$54.9&  4.69& 10.85&  1.84&  6.61& 14.25& 0.231& 0.497& 2.490&  98.9& 213.0 \\
  3& $-$25.7& $-$60.1&  4.66& 11.45&  1.63&  6.08& 12.94& 0.212& 0.452& 2.195&  70.8& 150.6 \\
  4& $-$51.1& $-$41.3&  4.65& 10.42&  1.64&  8.72& 16.72& 0.304& 0.583& 3.337& 102.7& 196.7 \\
  5&  $-$5.7&   1.7&  4.44&  8.91&  1.87& 10.79& 18.18& 0.377& 0.634& 5.578& 166.7& 280.9 \\
  6& $-$33.4& $-$49.0&  3.11& 10.62&  1.46&  8.44& 13.32& 0.295& 0.465& 1.655&  79.4& 125.4 \\
  7& $-$26.6&   4.6&  3.09&  9.64&  1.63&  7.60& 13.89& 0.265& 0.485& 1.616&  88.8& 162.1 \\
  8&   1.2&  34.7&  2.57&  9.16&  2.31&  8.67& 12.54& 0.303& 0.438& 1.582& 203.2& 293.7 \\
  9&  16.8&   5.3&  2.41& 10.11&  1.92&  9.88& 18.31& 0.345& 0.639& 3.271& 159.4& 295.7 \\
 10&   6.7& $-$32.0&  2.31& 11.13&  1.62&  6.04& 10.55& 0.211& 0.368& 0.969&  69.7& 121.6 \\
 11& $-$12.5&  22.7&  1.97&  8.05&  2.13&  5.36& 10.31& 0.187& 0.360& 0.565& 107.3& 206.2 \\
 12&   0.8& $-$21.0&  1.95&  9.61&  1.35&  9.15& 12.74& 0.319& 0.445& 0.727&  73.7& 102.5 \\
\noalign{\smallskip}
\hline
\end{tabular}
\smallskip\noindent
\\
\end{flushleft}
\end{table*}

\begin{table*}
\caption{Parameters of the clumps identified in the SEST CO(1--0) data}
\label{sestclumps}
\begin{flushleft}
\begin{tabular}{rrrcrcrrcccrr}
\hline
\noalign{\smallskip}
Clump & $\Delta\alpha$ & $\Delta\delta$ &$T_{\rm mb}$ & 
\multicolumn{1}{c}{$V_{\rm lsr}$} & $\Delta V$
 & \multicolumn{1}{c}{$r_1$} & \multicolumn{1}{c}{$r_2$} & 
$r_1$ & $r_2$ & $L_{\rm CO}$ & $M_{\rm vir}$($r_1$) & $M_{\rm vir}$($r_2$) \\  
      & \multicolumn{2}{c}{(arcmin)} & (K) & \multicolumn{2}{c}{(\kms)} & 
\multicolumn{2}{c}{(arcmin)} & \multicolumn{2}{c}{(pc)} & K\kms pc$^2$ & 
\multicolumn{2}{c}{(\Msol)} \\
\noalign{\smallskip}
\hline
\noalign{\smallskip}
   1& $-$0.6& $-$1.2& 4.70& 7.53& 0.57& 2.98& 6.11& 0.104& 0.213& 0.202& 4.3& 8.7 \\
   2& $-$18.9& $-$6.7& 4.56& 9.49& 0.73& 3.61& 7.63& 0.126& 0.266& 0.284& 8.4& 17.7 \\
   3& $-$5.0& 0.6& 4.37& 8.37& 0.94& 3.55& 7.44& 0.124& 0.260& 0.444& 13.7& 28.6 \\
   4& $-$2.9& 12.7& 4.27& 8.29& 0.94& 2.64& 4.99& 0.092& 0.174& 0.210& 10.2& 19.2 \\
   5& 6.1& 5.2& 3.71& 8.51& 0.68& 3.13& 6.47& 0.109& 0.226& 0.164& 6.3& 13.0 \\
   6& 0.7& 7.4& 3.53& 8.00& 0.83& 2.46& 4.87& 0.086& 0.170& 0.137& 7.5& 14.8 \\
   7& $-$23.7& $-$4.4& 3.36& 9.01& 0.59& 3.98& 5.52& 0.139& 0.193& 0.068&  6.2& 8.6 \\
   8& 5.1& $-$1.9& 3.42& 8.37& 0.99& 2.79& 5.58& 0.097& 0.195& 0.164& 11.9& 23.8 \\
   9& $-$11.4& $-$4.5& 3.29& 8.72& 0.85& 3.86& 7.81& 0.135& 0.273& 0.336& 12.1& 24.5 \\ 
  10& $-$25.3& $-$12.2& 3.51& 9.17& 0.69& 2.27& 4.45& 0.079& 0.155& 0.075& 4.7& 9.2 \\
  11& $-$20.6& $-$14.3& 3.69& 9.39& 0.50& 1.83& 3.44& 0.064& 0.120& 0.038& 2.0& 3.8 \\
  12& 14.3& 9.6& 3.56& 8.64& 0.85& 1.93& 3.83& 0.067& 0.134& 0.069& 6.2& 12.2 \\
  13& $-$25.3& 7.2& 3.03& 9.63& 1.13& 3.05& 6.25& 0.107& 0.218& 0.197& 17.3& 35.3 \\
  14& 18.3& 3.8& 2.76& 10.70& 0.95& 4.89& 9.54& 0.171& 0.333& 0.321& 19.4& 37.8 \\
  15& 13.1& $-$6.2& 2.74& 9.49& 0.80& 4.53& 6.42& 0.158& 0.224& 0.164& 12.6& 17.9 \\
  16& 25.7& 6.1& 2.73& 10.18& 0.90& 4.39& 7.92& 0.153& 0.277& 0.252& 15.5& 28.0 \\
  17& $-$13.0& $-$12.4& 2.78& 8.99& 0.59& 2.66& 4.75& 0.093& 0.166& 0.055& 4.1& 7.3 \\
  18& $-$6.8& $-$9.6& 2.73& 7.41& 0.90& 2.22& 3.83& 0.078& 0.134& 0.044& 7.9& 13.7 \\
  19& $-$6.6& 5.6& 2.55& 7.72& 0.83& 2.10& 4.45& 0.073& 0.155& 0.053& 6.4& 13.6 \\
  20& 8.5& 9.0& 2.18& 9.49& 0.95& 5.48& 10.37& 0.191& 0.362& 0.423& 21.5& 40.7 \\
  21& $-$32.1& 3.6& 2.26& 8.80& 0.49& 3.94& 5.63& 0.138& 0.196& 0.064& 4.1& 5.9 \\
  22& $-$9.5& 10.8& 2.27& 8.21& 1.21& 2.61& 3.27& 0.091& 0.114& 0.030& 16.7& 21.0 \\
  23& 15.0& 12.6& 2.20& 8.92& 0.67& 3.39& 6.06& 0.118& 0.212& 0.107& 6.8& 12.1 \\
  24& $-$31.0& $-$8.1& 2.30& 9.19& 0.66& 2.38& 3.76& 0.083& 0.131& 0.037& 4.5& 7.2 \\
\noalign{\smallskip}
\hline
\end{tabular}
\smallskip\noindent
\\
\end{flushleft}
\end{table*}

In Table~\ref{kosmaclumps} we give the following information for each clump:
a running number (Col.~1); offsets in Right Ascension and Declination (Cols.~2 
and 3) relative to the central position; 
peak temperature, velocity of the peak, and linewidth (Cols.~4--6); the 
HWHM-radius ($r_1$) in arcmin, derived from the dispersion in both 
coordinates, and corrected for beam size (Col.~7); the clump radius ($r_2$) 
in arcmin, derived from the total number of clump pixels above the adopted 
threshold value, assuming a circular distribution and corrected for beam size 
(Col.~8). In Cols.~9 and 10 we give $r_1$ and $r_2$, respectively, in parsec, 
assuming a distance of 120~pc. The CO luminosity, $L_{\rm CO}$ is in Col.~11, 
while the virial masses, assuming a density distribution $\rho \propto r^{-2}$ 
(i.e., $M_{\rm vir}({\rm M}_{\odot}) = 126 r \Delta V^2$), 
are given in Col.~12 ($r = r_1$) and Col.~13 ($r = r_2$).

\begin{figure*}
\sidecaption
\includegraphics[width=12cm]{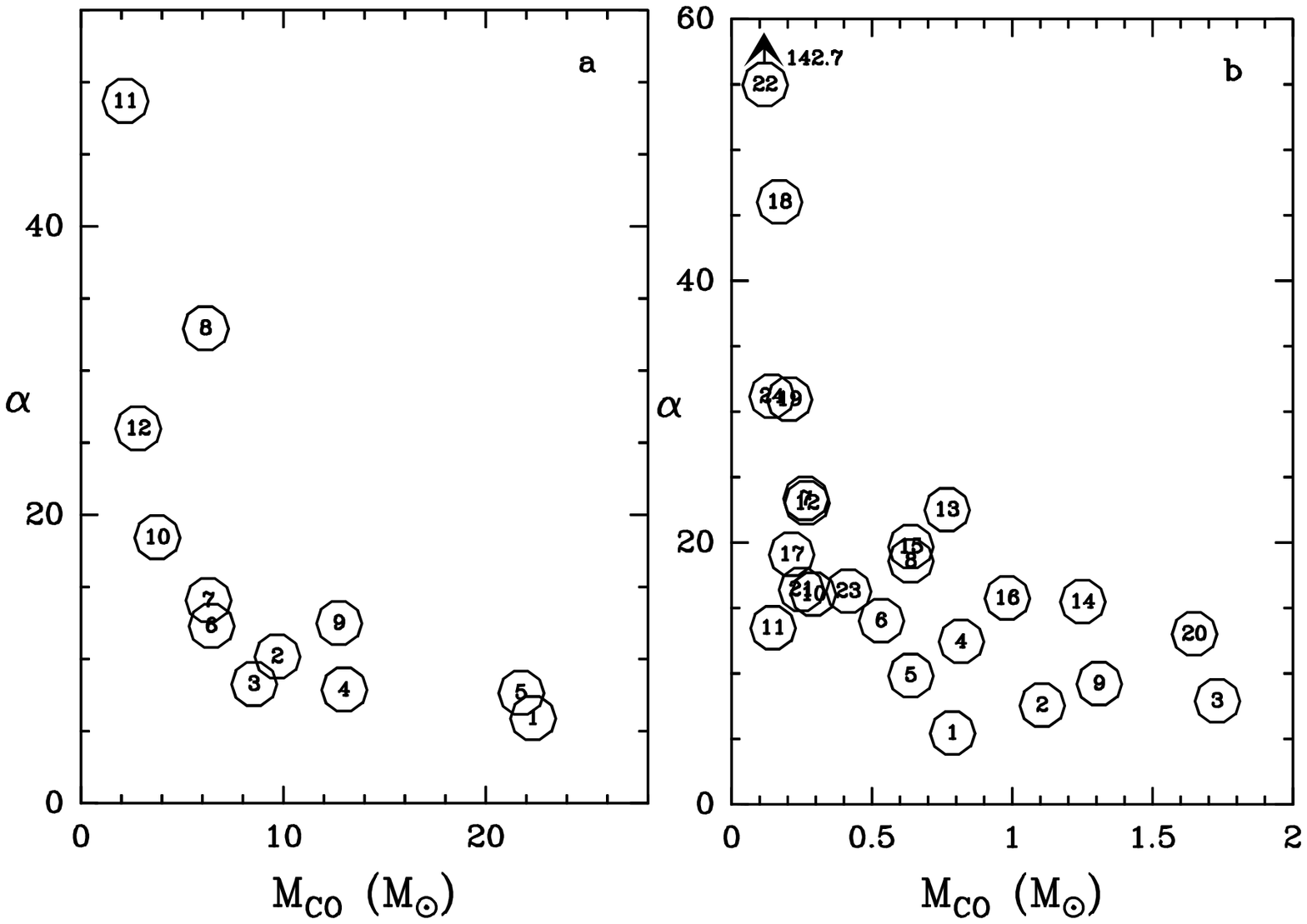}
\caption{
{\bf a.}\ Virial parameter $\alpha = M_{\rm vir}(r_1)/M_{\rm CO}$ as a function 
of $M_{\rm CO} = 3.9L_{\rm CO}$, for KOSMA-data. Data in Table~\ref{kosmaclumps}. 
The numbers in the circles identify the clumps.
In the absence of external pressure and magnetic fields, virial equilibrium 
is represented by $\alpha=1$.
{\bf b.}\ As a., but for SEST-data. Data in Table~\ref{sestclumps}.
For display purposes, clump 22 is not drawn at its real position, which is 
indicated with an arrow.
}
\label{virialparam}
\end{figure*}

Figure~\ref{virialparam}a shows the virial parameter 
$\alpha = M_{\rm vir}(r_1)/M_{\rm CO}$ as a function of 
$M_{\rm CO} = 3.9L_{\rm CO}$. The virial parameter is a measure of the extent 
to which a cloud or clump is in equilibrium. 
Under idealised conditions, $\alpha = 1$ 
for a clump in virial equilibrium, without external pressure and in the 
absence of a magnetic field; if the kinetic energy of a clump 
is balanced by its gravitational energy, $\alpha = 2$, and the clump is 
said to be gravitationally bound. In MBM~18 all clumps have $\alpha \ga 6$,
thus gravity is not the dominating force in any of these structures.

However, with observations performed with a 3\pam9 beam on a 4\amin\ raster, 
the 
KOSMA map is incompletely sampled, and this may be expected to influence the 
clump 
identification process. With the SEST we mapped only the northern part of the 
cloud, but at higher resolution (although still incompletely sampled). 
Running CLUMPFIND on these data (using 0.72~K
[=3$\sigma$] for the lowest contouring value and increment) results in 24 
clumps. The clumps' parameters are given in Table~\ref{sestclumps}, 
where the description of the columns is the same as for 
Table~\ref{kosmaclumps}. The total mass contained in the clumps (from 
their combined CO-luminosities) is 15.4~\Msol, which is only 27\% of the mass 
determined from all emission in the cloud (Sect.~\ref{sest}).
Apparently there is much diffuse emission that CLUMPFIND cannot assign to 
any clumps. For the SEST-clumps we plot the virial parameter $\alpha$ versus 
the clump mass $M_{\rm CO} = 3.9L_{\rm CO}$ in Fig.~\ref{virialparam}b. 
Also in these higher-resolution data we find that all clumps have 
$\alpha \ga 5.5$, implying that none of the clumps are gravitationally bound.

\subsection{CO pedestals}

\begin{figure}
\resizebox{9cm}{!}{\rotatebox{270}{\includegraphics{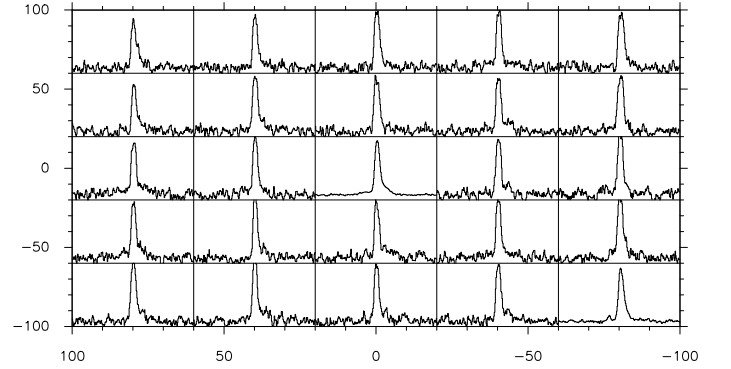}}}
\caption{Spectra of the $^{12}$CO(1--0) emission in the central 200\asec\ 
$\times$ 200\asec\ region of MBM~18 (SEST observations). Offsets are in 
arcseconds.
For each spectrum the velocity scale runs from $-5$ to $+20$~\kms, and the 
temperature scale (\TAstar) from $-0.5$ to 5.5~K. Note the non-Gaussian 
profiles at most positions.}
\label{coprofiles}
\end{figure}

\begin{figure}
\resizebox{9cm}{!}{\rotatebox{270}{\includegraphics{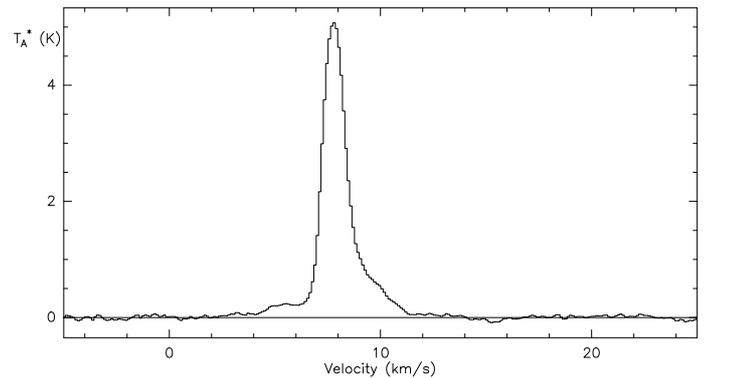}}}
\caption{Average $^{12}$CO(1--0) spectrum of the central 400\asec\ 
$\times$ 400\asec\ region of MBM~18 (SEST observations). The red and blue 
wings/pedestals are clearly visible.
}
\label{central400}
\end{figure}

Many of the SEST CO(1--0) profiles show evidence of red- and blue-shifted 
low-level emission "wings" (see Fig.~\ref{coprofiles}). This phenomenon is 
routinely associated with
bipolar outflows and is a hallmark of winds from an embedded YSO (e.g., 
Bachiller~\cite{bachiller}). However, this is not likely the case for MBM 18. 
No discernible blue- or red-shifted lobes can be traced out in the channel 
maps, nor can the winged profiles be associated distinctly with any of the
candidate T Tauri stars. It is more likely that the observed wings are 
a manifestation of the CO "pedestals" that are not associated with star
formation in translucent high-latitude clouds (Blitz et al.~\cite{blitz}; 
Magnani et al.~\cite{magnani90}). These pedestals are characterstic of the 
fragmented structure of these clouds, a likely consequence of turbulence.
Shore et al. (\cite{shore03}) and Barriault et al. (\cite{barriault}) proposed 
that high-latitude 
clouds may form at the confluence of atomic gas flows. In these 
scenarios, a turbulent cascade smears out large-scale velocity gradients, 
leaving only small-scale centroid fluctuations (Shore et al.~\cite{shore06}).
Observationally, these small-scale velocity fluctuations manifest themselves
as low-level emission at velocities other than the cloud average velocity
where the bulk of the emission lies. Figure~\ref{central400} is the 
average spectrum of the central 400\asec\ $\times$ 400\asec\ 
region observed in CO(1--0) with SEST (Fig.~\ref{sest_tp_ti}). The bulk of the 
CO emission is concentrated in a Gaussian profile with \Vlsr $\sim 8$~\kms. 
The wings or pedestals
are clearly noticeable, extending from 4 to 11~\kms. We note that this 
phenomenon is not intermittency as described by Falgarone and Phillips 
(\cite{falgarone}), which one would
expect to be much weaker and not visible in individual spectra like those 
shown in Fig.~\ref{coprofiles}. To detect the signature of intermittency, a 
probability density function (PDF) analysis should be performed, which would 
require extending
our high-resolution observations over the entire cloud. In summary, the lack
of any clear bipolar structure to the wing emission, and the absence of
correlation of the wings with the positions of the candidate T Tauri stars 
indicate that the wings seen in
individual and composite profiles of the central region in MBM~18 are not 
evidence of star formation but, rather, of the fragmented structure of the
molecular cloud as a consequence of turbulence.

\section{Conclusions}

The main objective of the observations presented and analysed in this paper, 
is to investigate whether low-mass star formation occurs in the translucent 
cloud MBM~18 (LDN~1569). 

We have mapped the $^{12}$CO(1--0) emission in the high-latitude cloud MBM~18 
(LDN~1569) with the KOSMA 3-m telescope; this is the first time this object 
has been fully mapped in CO. The northern part of the cloud was 
also mapped at higher resolution in $^{12}$CO(1--0) and $^{13}$CO(1--0) with 
the SEST.
The total cloud mass, estimated from the integrated CO emission, is 162~\Msol, 
assuming a distance of 120~pc and $X = N(H_2)$/\ITmb\ = $1.8 \times 
10^{20}$~cm$^{-2}$(Kkm\,s$^{-1}$)$^{-1}$. 
The mapping data show that the cloud contains substantial spatial 
substructure, while the channel maps also reveal significant velocity 
structure. The virial mass, as determined from 
the standard deviation of the central velocities of Gaussian fits to the 
emission profiles at all detected positions, is at least 10$^3$~\Msol, 
and thus much larger than the gravitational mass. This implies that 
MBM~18 is not gravitationally bound. 
We used CLUMPFIND to analyse these substructures and found 12 clumps 
(KOSMA data), with beam-corrected equivalent (HWHM) radii ranging from 
0.19~pc to 0.38~pc, molecular masses between 2.2~\Msol\ and 22~\Msol, and 
virial masses from 70~\Msol\ to 203~\Msol. All these clumps 
have a virial parameter $\alpha = M_{\rm vir}/M_{\rm CO} \ga 6$ and are are 
consequently not gravitationally bound.
It is unlikely that stars could have formed in these clumps.
Clumps found in the SEST map have smaller masses, but show a similar 
picture, with $\alpha \ga 5.5$.

Inside the molecular boundaries of MBM~18 we found ten candidate H$\alpha$ 
emission-line stars from a low-resolution objective grism survey. To 
determine whether these T Tauri star candidates are indeed pre-main-sequence 
stars, we obtained higher-resolution spectra of seven of these 
stars to determine their spectral types, verify the presence of H$\alpha$ 
emission and to try and detect the presence of LiI absorption. One of the 
observed stars (Ha4) has the spectrum of a late F -- early G-type star, with 
H$\alpha$ in absorption.
Its deviating position in the near- and mid-IR 
colour-colour diagrams may be caused by a companion that is affecting the 
colours, but remains undetected in the spectrum. 
The other six stars are M-type dwarfs of type M1.5$-$M4.0 (see 
Table~\ref{tngresults}). Only four out of six have H$\alpha$ 
emission, none has other emission lines, and no strong (or indeed any)  
LiI absorption (only three stars show a mere hint) has been detected, 
implying that they are not T Tauri stars. From the observed colours 
and the intrinsic colours determined from the spectral types, we derive 
distances to the six M-stars of between 60 and 250~pc. 
We did obtain a good fit of the late-type stars on pre-main-sequence 
isochrones, finding ages between 7.5 and 15~Myr. By this time any lithium 
is likely to have been depleted. 
Only three of our target stars (viz Ha1, 2, and 8) lie 
near or within the generally accepted range of distances ($120-150$~pc) of 
MBM~18. 
If these stars were indeed formed in the cloud, 
MBM~18 would be considerably older than 
a typical translucent cloud. Considering that only unbound 
clumps are found, a formation {\it in situ} is unlikely. 

\begin{acknowledgements}
This work is partly based on observations made with the Italian
Telescopio Nazionale Galileo (TNG) operated on the island of La Palma by the
Fundaci\'on Galileo Galilei of the INAF (Istituto Nazionale di Astrofisica) 
at the Spanish Observatorio del Roque de los Muchachos of the Instituto de 
Astrofisica de Canarias (programmes TAC34-AOT19 and TAC17-AOT22).
The TNG-observations were carried out in service mode. We are grateful to our 
support astronomers Vania Lorenzi (2009 observations) and Francesca 
Ghinassi (2010 observations) for their assistance. We thank Giovanna Stirpe 
for help with the TNG-data reduction.\hfill\break\noindent
At the time of the observations, 
the KOSMA radio telescope at Gornergrat-S\"ud Observatory was operated by the 
University of K\"oln, and supported by the Deutsche Forschungsgemeinschaft 
through grant SFB-301, as well as by special funding from the Land 
Nordrhein-Westfalen. The observatory was administered by the Internationale 
Stiftung Hochalpine Forschungsstationen Jungfraujoch und Gornergrat, Bern, 
Switzerland. The telescope is currently located in Tibet.\hfill\break\noindent
This research has made use of the SIMBAD database, operated at CDS, Strasbourg,
France, and of NASA's Astrophysics Data System Bibliographic Services (ADS).
This publication makes use of data products from the Wide-field Infrared Survey 
Explorer, which is a joint project of the University of California, Los Angeles, 
and the Jet Propulsion Laboratory/California Institute of Technology, funded by 
the National Aeronautics and Space Administration.\hfill\break\noindent
We made use of observations with the 100-m telescope of the Max-Planck-Institut 
f\"ur Radioastronomie at Effelsberg.\hfill\break\noindent
We thank Adam Schneider for help in producing Fig.~\ref{wise}, and Steven 
Shore for comments on an earlier version of the paper.
\end{acknowledgements}


\begin{appendix}

\section{Spectra of the candidate H$\alpha$ emission-line stars}

In Fig.~A1 we present the full spectra of the M-dwarf stars in our sample. 
In Fig.~A2 we show the spectrum of the earlier type star.

\begin{figure*}
\resizebox{17cm}{!}{\rotatebox{270}{\includegraphics{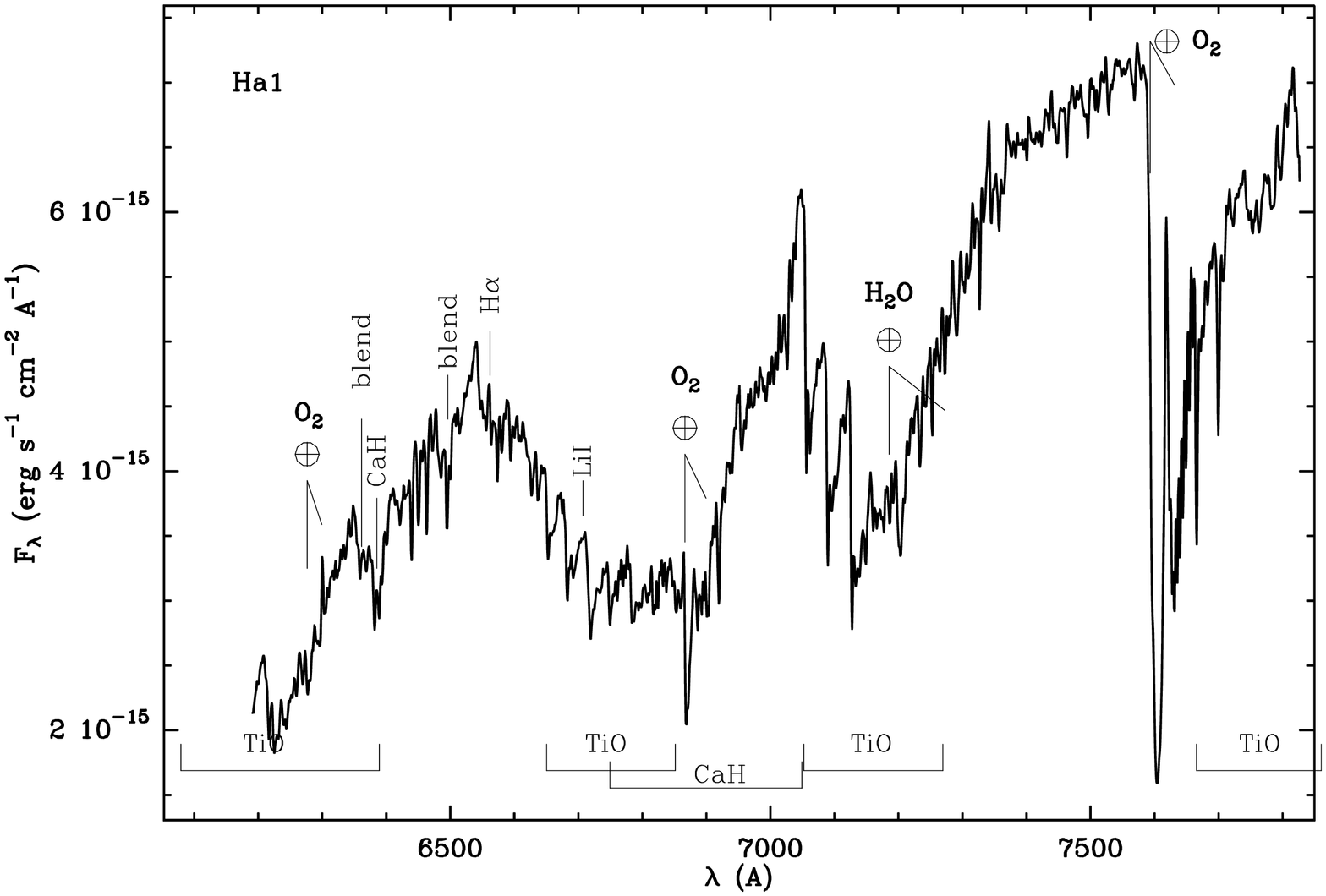}}}
\resizebox{17cm}{!}{\rotatebox{270}{\includegraphics{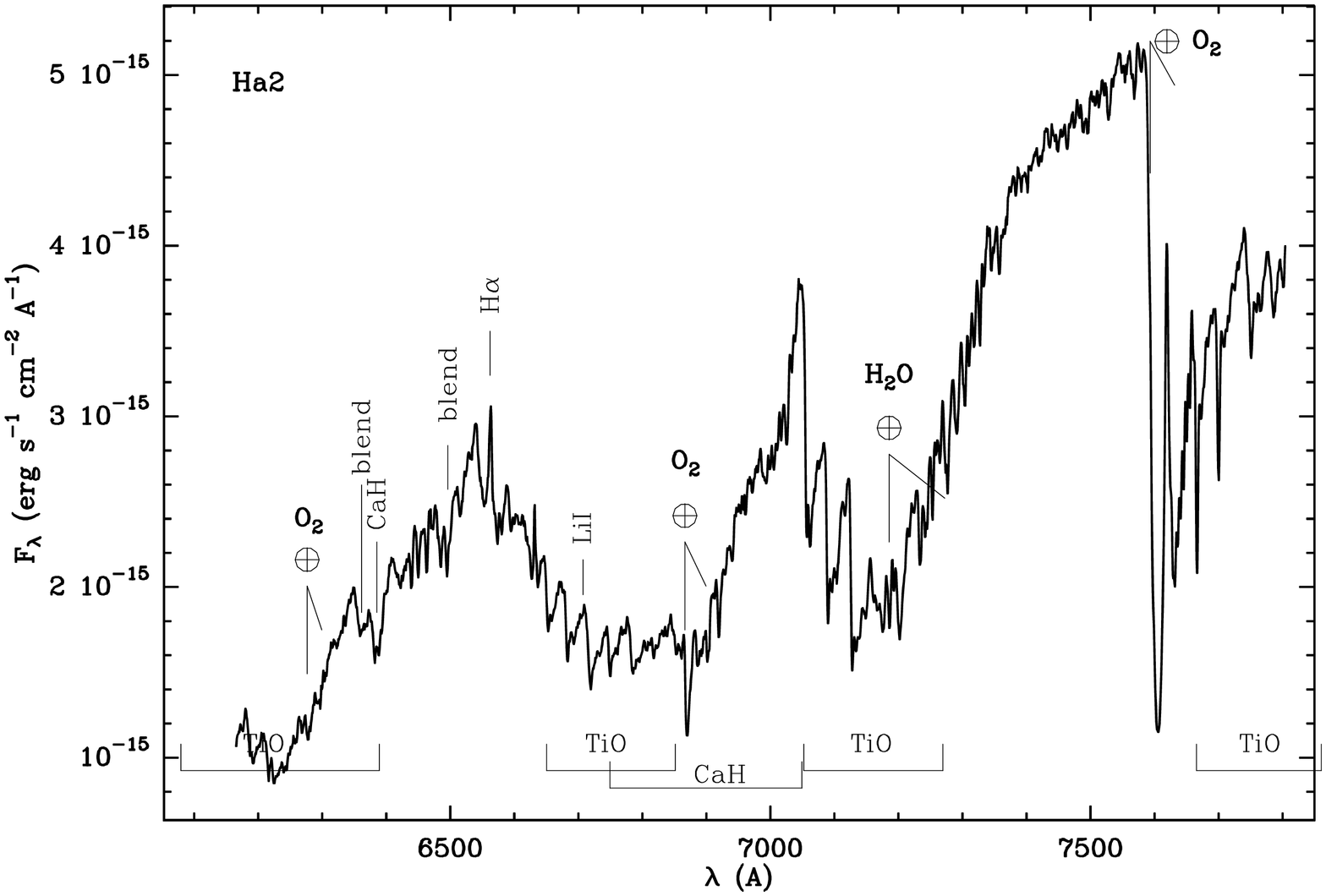}}}
\caption{
{\bf a.}\ (top panel) Spectrum of candidate H$\alpha$ emission-line star Ha1. 
The locations of the TiO and CaH absorption bands are indicated below the 
spectrum, as are the expected locations of several features. 
Telluric lines are marked with a crossed circle.
{\bf b.}\ (bottom panel) as a., but for candidate H$\alpha$ emission-line 
star Ha2.}
\label{fullspectra1}
\end{figure*}

\addtocounter{figure}{-1}

\begin{figure*}
\resizebox{17cm}{!}{\rotatebox{270}{\includegraphics{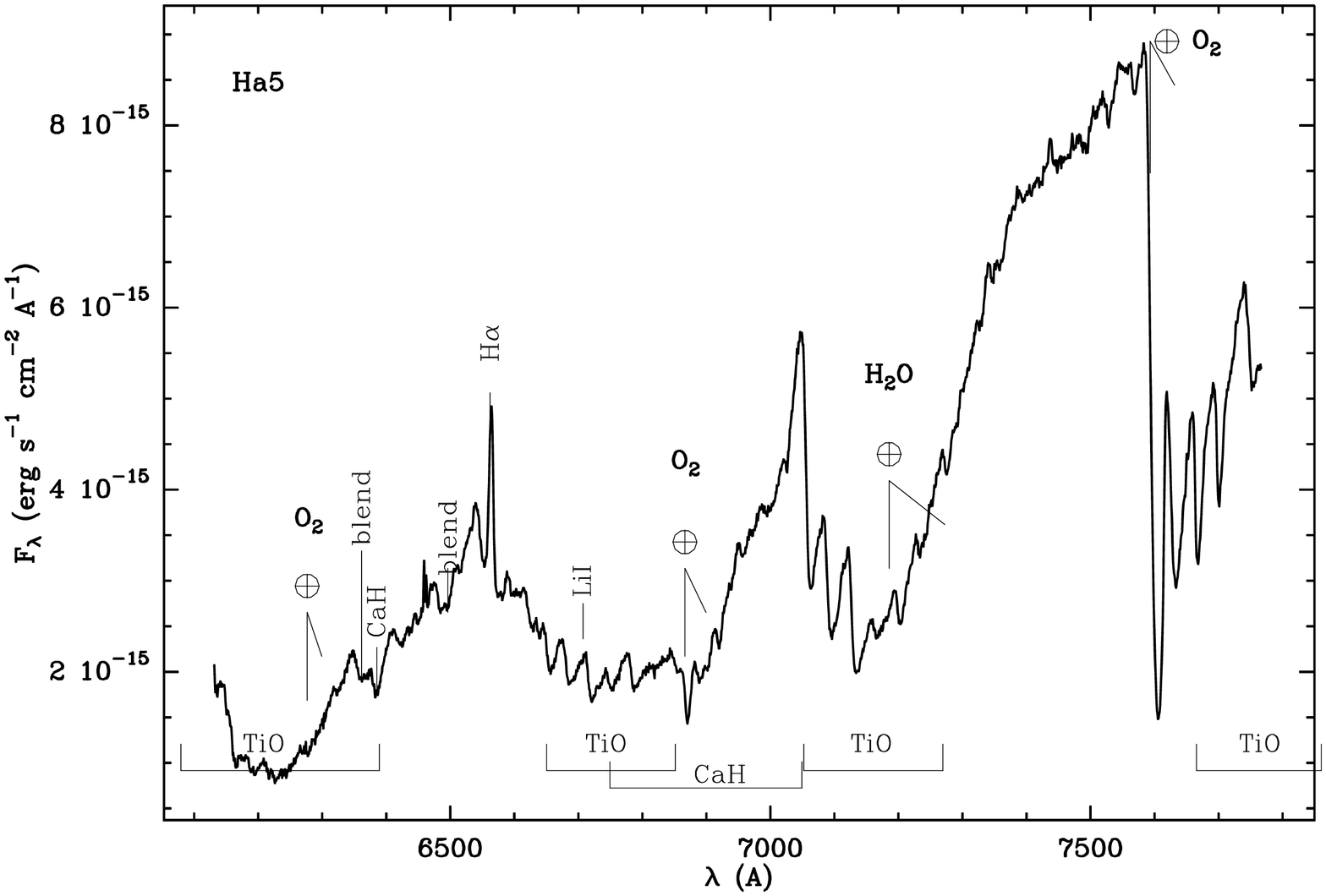}}}
\resizebox{17cm}{!}{\rotatebox{270}{\includegraphics{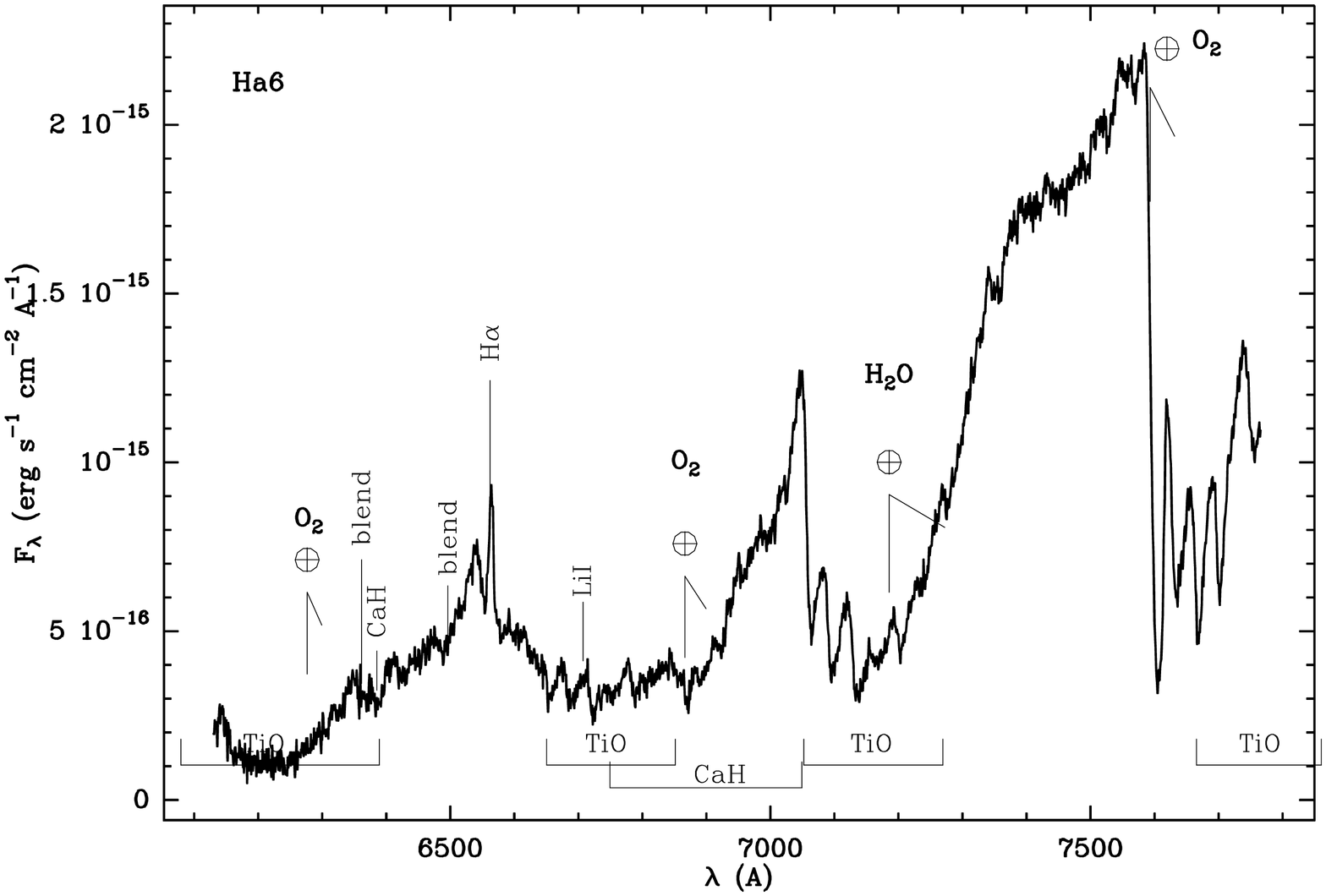}}}
\caption{
{\bf c.}\ (top panel) As Fig.~\ref{fullspectra1}a, for candidate 
H$\alpha$ emission-line star Ha5. 
{\bf d.}\ (bottom panel) as c., but for candidate H$\alpha$ emission-line 
star Ha6.}
\label{fullspectra2}
\end{figure*}

\addtocounter{figure}{-1}

\begin{figure*}
\resizebox{17cm}{!}{\rotatebox{270}{\includegraphics{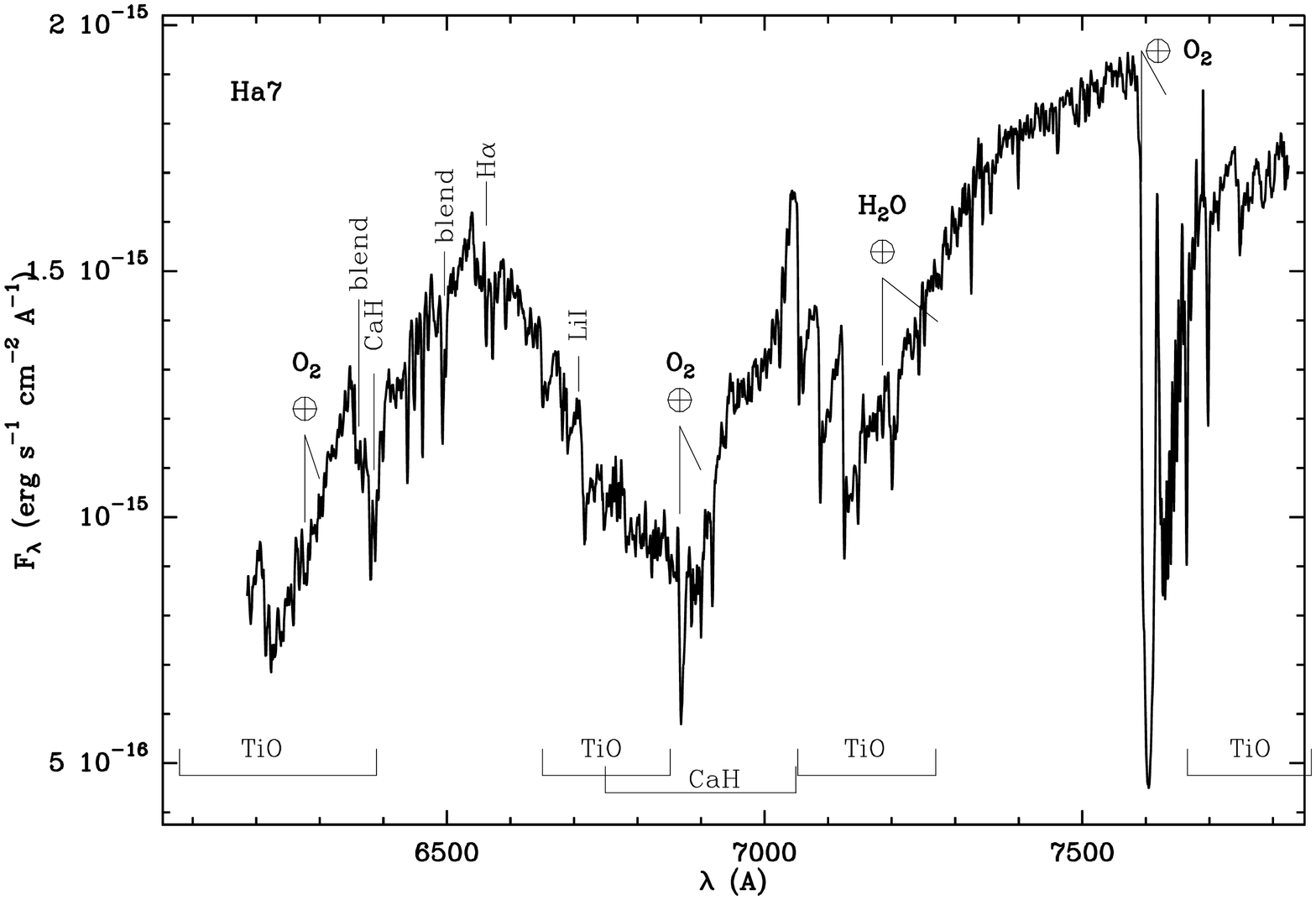}}}
\resizebox{17cm}{!}{\rotatebox{270}{\includegraphics{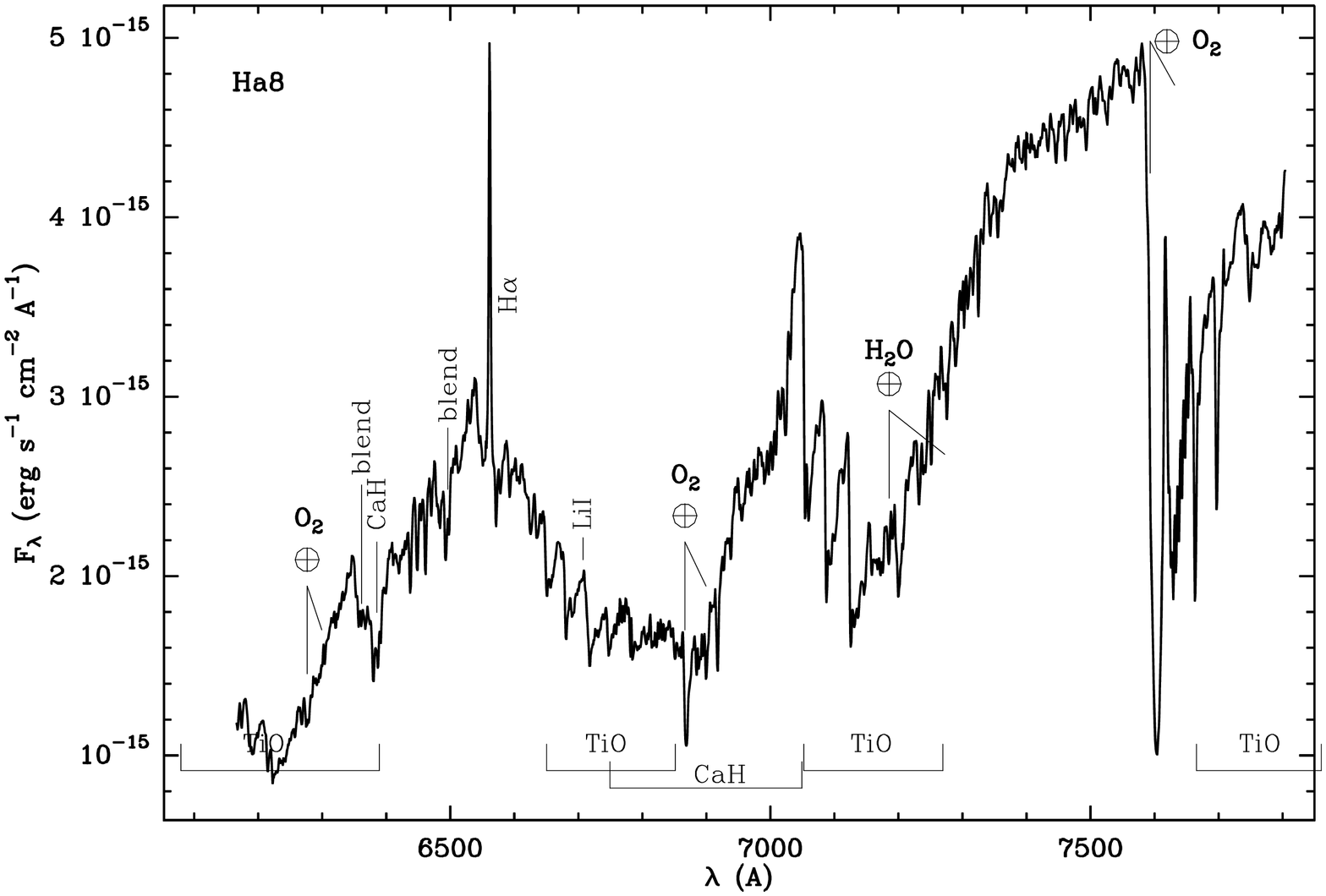}}}
\caption{
{\bf e.}\ (top panel) As Fig.~\ref{fullspectra1}a, for candidate 
H$\alpha$ emission-line star Ha7.
{\bf f.}\ (bottom panel) as e., but for candidate H$\alpha$ emission-line 
star Ha8.}
\label{fullspectra3}
\end{figure*}

\begin{figure*}
\resizebox{17cm}{!}{\rotatebox{270}{\includegraphics{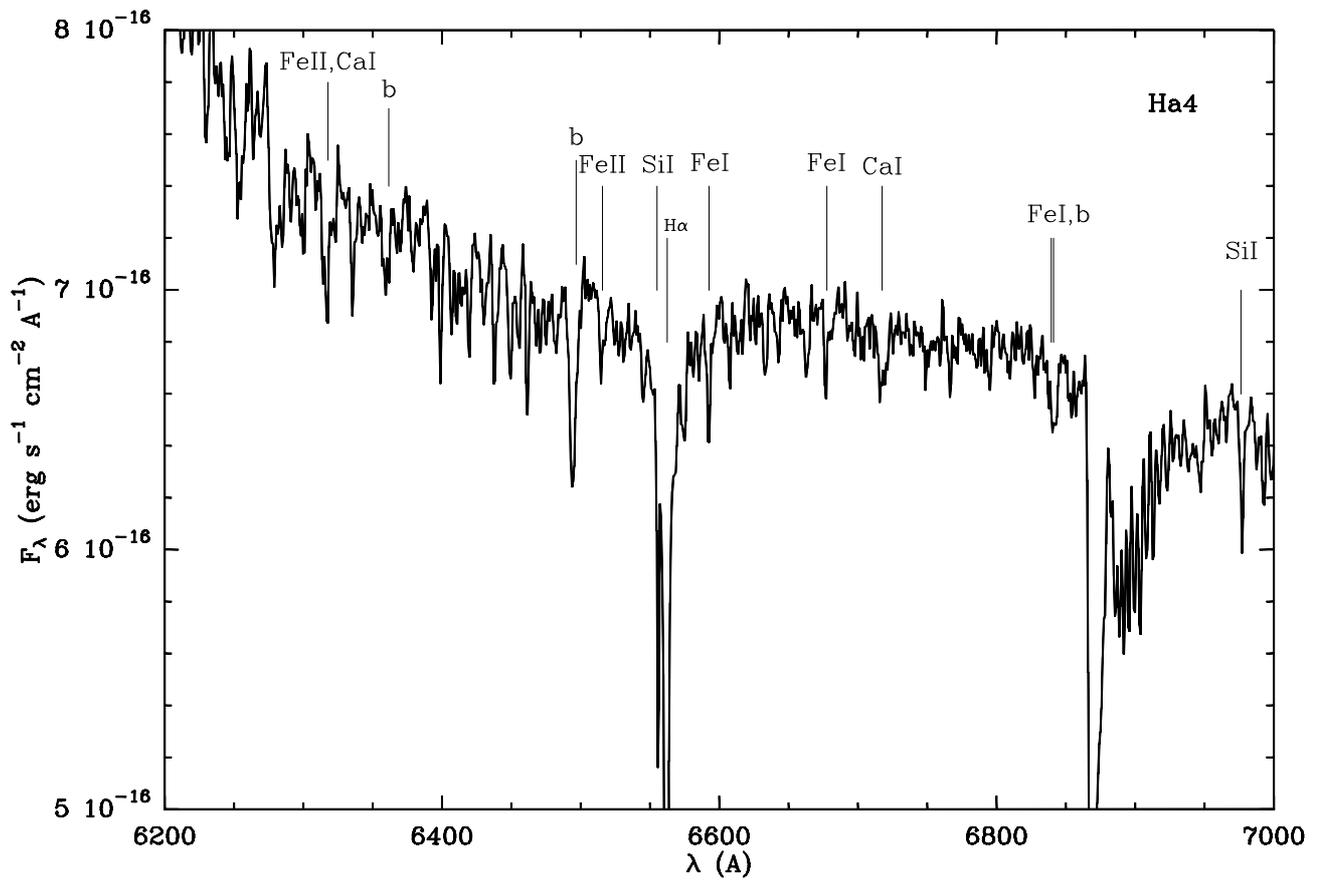}}}
\caption{
The spectrum of candidate H$\alpha$ emission-line star Ha4. Several 
absorption lines are indicated. This star is clearly of earlier spectral 
type than the others.}
\label{fullspectra4}
\end{figure*}

\end{appendix}

\listofobjects
\end{document}